\documentclass[preprint2,times,tighten]{aastex6}

\usepackage{color}
\usepackage{enumitem}
\usepackage{hyperref}
\usepackage{verbatim}
\usepackage{amsmath,amstext,mathrsfs}
\usepackage[all]{hypcap} 
\usepackage{afterpage}   
\usepackage{marginnote}

\begin{document}
\title{Distribution of velocity {\torefereetwo gradient} {\toreferee orientations}: \\Mapping magnetization with the Velocity Gradient Technique}
\author{A. Lazarian\altaffilmark{1}, Ka Ho Yuen\altaffilmark{1}, Ka Wai Ho\altaffilmark{2}, Junda Chen\altaffilmark{1}, Victor Lazarian\altaffilmark{1}, Zekun Lu\altaffilmark{1,3},Bo Yang\altaffilmark{1}, Yue Hu\altaffilmark{4}}
\email{lazarian@astro.wisc.edu }
\altaffiltext{1}{Department of Astronomy, University of Wisconsin-Madison}
\altaffiltext{2}{Department of Physics, The Chinese University of Hong Kong}
\altaffiltext{3}{School of Astronomy and Space Science, Nanjing University, Nanjing, China}
\altaffiltext{4}{College of Electronics and Information Engineering, Tongji University, Shanghai, China}

\begin{abstract}
Recent developments of the Velocity Gradient Technique (VGT) show that the velocity gradients provide a reliable tracing of magnetic field direction in turbulent plasmas. In this paper, we explore the ability of velocity gradients to measure the magnetization of interstellar medium. We demonstrate that the distribution of velocity gradient orientations provides a reliable estimation of the magnetization of the media. In particular,  we determine the relation between Alfvenic Mach number $M_A$ {\toreferee in the range of $M_A \in [0.2,1.7]$} and properties of the velocity gradient distribution, namely, with the dispersion of velocity gradient orientation as well as with the peak to base ratio of the amplitudes. We apply our technique for a selected GALFA-HI region and find the results consistent with the expected behavior of $M_A$. Using 3D MHD simulations we successfully compare the results with our new measure of magnetization that is based on the dispersion of starlight polarization. We demonstrate that, combined with the velocity dispersion along the line of sight direction, our technique is {\toreferee capable to delivering} the magnetic field strength. The new technique opens a way to measure magnetization using other  gradient measures such as synchrotron intensity gradients (SIGs) and synchrotron polarization gradients (SPGs). 
\end{abstract}

\section{Introduction}

Diffuse interstellar media and molecular clouds are both turbulent and magnetized \citep{1981MNRAS.194..809L,2004ARA&A..42..211E,2004RvMP...76..125M,2007prpl.conf...63B,MO07,CL09}. Magnetization of the media is extremely important for understanding key astrophysical problems, e.g.
the problem of star formation \citep{1959ApJ...129..243S,1956MNRAS.116..503M,1976ApJ...210..326M,1977ApJ...214..488S,1992pavi.book.....S,1991ApJ...370L..39K,1994ApJ...429..781S,1998ApJ...498..541K,1998ARA&A..36..189K,2013ApJ...770..151C,2013ApJ...768..159H,2015ApJ...808...48B,2018arXiv180105428B} and cosmic ray propagation and acceleration \citep{1949PhRv...75.1169F,1964ocr..book.....G,1966ApJ...146..480J,1978MNRAS.182..147B,1965ApJ...142..584P,1979cmft.book.....P,2014ApJ...783...91C}.

Various techniques have been proposed to study the magnetization of the interstellar media \citep{2009ASPC..414..453D}.  Magnetic field strength and structures can be measured by both direct or indirect approaches. Spectral line splitting due to Zeeman effect provides the direct measurement of the line of sight component of magnetic field strength \citep{1989ApJ...338L..61G,1993prpl.conf..279H,1999ApJ...520..706C,1999ApJ...514L.121C,2001ApJ...554..916B,C10,2008A&A...487..247F}, but the method {\toreferee is mostly applied to lower Galactic latitude target (Crutcher 2012)} and does not provide an insight on the total magnetization nor the magnetic field directions perpendicular to the line of sight {\toreferee unless the two circular polarization components are both resolved (Crutcher 2007)}. Polarized dust emission \citep{2012ApJS..201...13V,2000PASP..112.1215H,1982MNRAS.200.1169C,1986ApJ...308..270D,1997ApJ...487..320N,1999MNRAS.303..659H,1984ApJ...284L..51H,1988QJRAS..29..327H} as well as measuring of starlight polarization arising {\toreferee from } aligned dust \citep{2012ApJS..200...19C,2012ApJS..200...21C,2013AJ....145...74C,Aetal15} provides the main indirect method of mapping the plane-of-sky magnetic field strength using the {\toreferee Davis-Chandrasekhar-Fermi }\citep{D51,CF53,Fal08,2009ApJ...696..567H,2009ApJ...706.1504H,2011ApJ...733..109H,2016ApJ...820...38H,2012ApJ...749...45C} method , but the estimation of magnetic field strength using the CF method can be significantly different from the true value even in synthetic maps obtained with MHD simulations for the case of trans-Alfvenic turbulence \citep{Fal08, CY16}.

Anisotropies of magnetized turbulence (see \citealt{BL13} for a review of the theory) provides an alternative way to study magnetic field directions. \cite{Letal02}  demonstrated that using the correlation functions of intensities {\toreferee in} velocity channel maps\footnote{Intensity fluctuations {\toreferee within "thin" (See \S \ref{sec:numerics})} velocity channel maps are in most cases dominated by velocity fluctuations \cite{LP00, LP04}},{\toreferee it is feasible} to study the magnetic field directions. Further development of the technique showed that the the Correlation Functions Anisotropy (CFA, see also Appendix \S \ref{subsec:cfa}) is a promising way to study both magnetic field direction and media magnetization, as well as to distinguish the contributions from Alfven, slow and fast modes (see \citealt{Eetal15,KLP16,KLP17a}). The CFA technique shows effective results on both the channel maps and velocity centroids \citep{EL05, KLP17b}. The CFA is also applicable to the study magnetic fields and turbulence properties with synchrotron intensity and polarization fluctuations \citep{LP12, LP16}. 

The approach in \cite {Letal02} can be realized in different ways with CFA being its particular realization. For instance, the anisotropies can be studied by employing the Principal Component Analysis (PCA) technique as described in \cite{Hetal08}. Our study in \cite{brazil18} showed, however, that there are no particular advantages of this technique compared to either VGT or CFA. Therefore we do not discuss this technique further in this paper. 

A different approach of magnetic field tracing with gradients of observable measures (e.g. velocity centroids, synchrotron intensity and synchrotron polarization, etc.) has been suggested recently \citep{GCL17, YL17a, YL17b, LY18a, LY18b, Letal17}. This technique in terms of theoretical justification is related to the CFA. {\toreferee It is known that the MHD fluctuations can be decomposed into fundamental Alfven, fast and slow modes (Biskamp 2003, {\torefereetwo See Appendix \ref{App_A}}). A numerical study in Cho \& Lazarian (2003) testifies that the three modes are evolving creating their own cascades with Alfven and slow modes showing significant anisotropy along the magnetic  field direction. With the contribution of Alfven and slow modes being dominant, especially in the weakly compressible cases (Lithwick \& Goldreich 2001, Cho \& Lazarian 2002), {\torefereetwo a prominent anisotropy is both expected and observed in the MHD turbulence containing all the three MHD modes }(see Cho \& Lazarian 2003, Kowal \& Lazarian 2009).} The corresponding  elongation of turbulent fluctuations causes the correlations of velocity to be stronger along the local magnetic field direction. At the same time the gradients of velocity {\toreferee become} perpendicular to magnetic field. As a result, one can estimate the direction of magnetic field by the direction of velocity gradients through a 90-degree rotation on the local gradient direction. Similarly, {\toreferee one can infer} the direction of magnetic fields {\toreferee by rotating the the magnetic gradients by 90 degrees} . The latter can be revealed through studying synchrotron intensity and synchrotron polarization gradients. For the sake of simplicity, within this study we shall focus on velocity gradients. However in view of the symmetric way how velocity and magnetic fluctuations enter Alfvenic turbulence, all our results in this paper are applicable also to magnetic field gradients.

Density fluctuations for low sonic Mach number $M_s$ will follow the velocity statistics. Thus for low $M_s$, density gradients behave similarly to velocity gradients. However, density is not always a passive scalar of velocity information and the density gradients created by shocks tend to be parallel to magnetic field in the case of high $M_s$. Therefore a combination of density and velocity gradients provides a better insight into the properties of turbulence in diffuse media. 

In Alfvenic turbulence the directions of ${\bf k}$ vectors of turbulent velocities has a statistical distribution (see GS95,\citealt{2002ApJ...564..291C}). Similarly, the gradients also have a distribution of directions with the most {\toreferee probable} orientation of gradients being perpendicular to the magnetic field direction. To find this most probably orientation, in practical studies, the velocity gradients are calculated over a block of data points.  When the statistics is sufficiently rich, the histogram of gradient orientations within a block gets Gaussian with the peak of the Gaussian corresponding to the local direction of magnetic field within the block.\footnote{This procedure should not be confused with the Histogram of Relative Orientations (HRO) technique explored by \cite{Setal13} for the intensity gradients. The latter requires polarimetry data to define the direction of magnetic field and draws the {\it relative} orientation of polarization directions and intensity gradients as a function of column density. Our technique is polarization-independent and is the way of finding the magnetic field direction, which is different from the purpose of the HRO. We stress that our approach when applied to velocities and densities provides the spacial direction of magnetic field, while the HRO provides the correlation of the relative orientation of intensity gradients as function of column density {\toreferee (see more comparison in YL17b, LY18a)}.}

In our earlier studies (e.g. \citealt{YL17a,LY18a}) we demonstrated that the peak of the distribution is correlated with the magnetic field direction and this provides a promising way of magnetic field studies, including studies of the 3D distribution of magnetic fields.\footnote{The 3D studies are possible with the {\toreferee Velocity Channel Gradients (VChGs)} using the galactic rotation curve to distinguish different emitting regions \citep{LY18a}. Another way of obtained 3D magnetic field structure employs the Faraday depolarization within the synchrotron polarization gradient studies \citep{LY18b}.} Our earlier studies were focused on magnetic field tracing by determining the peak of the Gaussian distribution of gradients (see \citealt{YL17a, YL17b, Letal17, LY18a, LY18b}). However, it became clear that the distribution of gradient orientations\footnote{\toreferee Similar to polarization, the gradients have $180^o$ ambiguity in determining the magnetic field actual direction. } is also an informative measure. For instance, the dispersion of the gradient orientation distribution was used to identify the regions of collapse induced by self-gravity \citep{LY18a}.\footnote{In the regions of self-gravitational collapse, the relative direction of velocity gradients and magnetic field changes gradually from perpendicular to parallel. This induces an increase of the dispersion \citep{YL17b, LY18a}.} However, the gravitational collapse is a special case when the properties of the turbulent flow change dramatically. In this paper we focus our attention on the properties of the distribution of gradient orientations for magnetized turbulence in diffuse regions where the effects of self-gravity is negligible. We are going to show that for such settings the properties of the distribution of gradient orientations are directly related to the Alfven Mach number $M_A$, which is the ratio of the turbulent velocity $V_L$ at the scale $L_{inj}$ and the media Alfven velocity $V_A$. We shall also demonstrate the ways of using spectroscopic data in order to convert the distribution of $M_A$ into the distribution of magnetic field strengths. 

In a companion paper by \cite{YLL18} we found that the amplitude of velocity gradients can be used to study both sonic $M_s=V_L/c_s$, where $c_s$ is the sound speed. Thus two most important measures of turbulence, $M_A$ and $M_s$ can be obtained using the gradient techniques.

In this paper, we focused on the relation of velocity gradient dispersion to $M_A$, and thus provide a way to map the magnetization in the media. In what follows, we provide the theoretical justification of this work in \S \ref{sec:theory}, discuss numerical simulations that we employ to test our expectations in \S \ref{sec:numerics}, and analyze our results in \S \ref{sec:result}, including an application to observation in \S \ref{sec:observation}. We briefly discuss the ways of obtaining magnetic field strength using our estimation of $M_A$ in \S \ref{intensity}. Our discussion and summary are in \S \ref{sec:discussion} and \S \ref{sec:summary}, respectively.

\section{Theoretical considerations}
\label{sec:theory}

In what follows we consider fluctuations of velocity arising from MHD turbulence. Similar considerations, however, are applicable to the fluctuations of turbulent magnetic field.

\subsection{Basic MHD turbulence considerations}

The predictive theory of incompressible MHD turbulence was formulated in \cite{GS95}, henceforth GS95. This theory can be understood on the basis of the Kolmogorov hydrodynamic turbulence theory if a concept of fast turbulent reconnection \citep{LV99}, henceforth LV99 is added. Indeed, according to LV99, magnetic reconnection happens in just one eddy turnover time for eddies at all scales. {\toreferee Therefore,  motions that mix magnetic field lines perpendicular to their direction is the way at which the turbulent cascade does not need to bend magnetic field lines and therefore encounters the least resistance from the field.} As a result, the turbulent eddies are elongated along the direction of magnetic field. Incidentally, this justifies the concept of {\toreferee alignment of turbulent eddies} with the local magnetic field direction that was not a part of the original GS95 picture\footnote{The closure relations employed in GS95 assume that the calculations were done in the reference frame of the mean field. Numerical simulations confirm that the GS95 relations are not correct in the mean field reference frame, but only correct in the local magnetic field reference frame.} but the necessity of using the local frame of reference was pointed out by numerical simulations (see \citealt{CV00, MG01}). 

The modern theory of MHD turbulence is discussed in e.g. \cite{BL13}. Here we briefly summarize the points that are essential for understanding the properties of gradients (see also Appendix \ref{App_A}). {\toreferee As we explain in the Appendix \ref{App_A} the properties of Alfvenic modes provide the basis for the gradient techniques. These modes also imprint their structure on the slow modes, while the fast modes provide a subdominat contribution. Thus we focus below on Alfvenic turbulence. }

Assume that the injection of turbulent energy takes place at the scale $L_{inj}$. If injection velocity $V_L$ equals to the Alfven velocity $V_A$, the Alfvenic turbulence naturally produces the Kolmogorov scaling in the direction perpendicular to magnetic field. Indeed, as we mentioned earlier, motions induced by Alfvenic turbulence are not constrained by magnetic field and therefore they produces a cascade of energy $ v_l^2$, where  $l$  is the eddy size perpendicular to magnetic field. The cascading to smaller scales happens over the eddy turnover time is $\sim l/v_l$. For $l$ significantly larger than the dissipation scale, the flow of kinetic energy $v_l^3/l$ is constant. This way one easily gets the GS95 scaling for the motions perpendicular to the magnetic field, i.e. $v_l\sim l^{1/3}$. 

To find how eddies evolve in the direction parallel to magnetic field one should take into account that the mixing motions associated with magnetic eddies send Alfven waves with the period equal to the period of an eddy, i.e. 
\begin{equation}
l/v_l\approx l_{\|}/V_A,
\label{critbal}
\end{equation}
where $l_{\|}$ is the parallel scale of the eddy. The {\toreferee condition (Eq. \ref{critbal})} is associated in GS95 theory with the critical balance {\toreferee stressing the fact that Alfven modes, which is by definition incompressible (See Biskamp 2003), have zero velocity divergence all over the space. As a result, the infall velocity gradient $v_l/l$ should be equivalent to that propagating velocity gradients $V_A/l_{\|}$ of the Alfvenic wave along the magnetic field line}. Combining this with the relation for $v_l\sim l^{1/3}$  one gets relation between the parallel and perpendicular scales of the eddies, i.e. $l_{\|}\sim l^{2/3}$. This is the relation that is true for the eddies that are aligned with the {\it local} direction of the magnetic field that surrounds them. The velocities associated with a turbulent eddy are anisotropic,{\toreferee and thus} the largest change of the velocity is in the direction perpendicular to the local direction of magnetic field. 

\begin{figure}[t]
\centering
\includegraphics[width=0.49\textwidth]{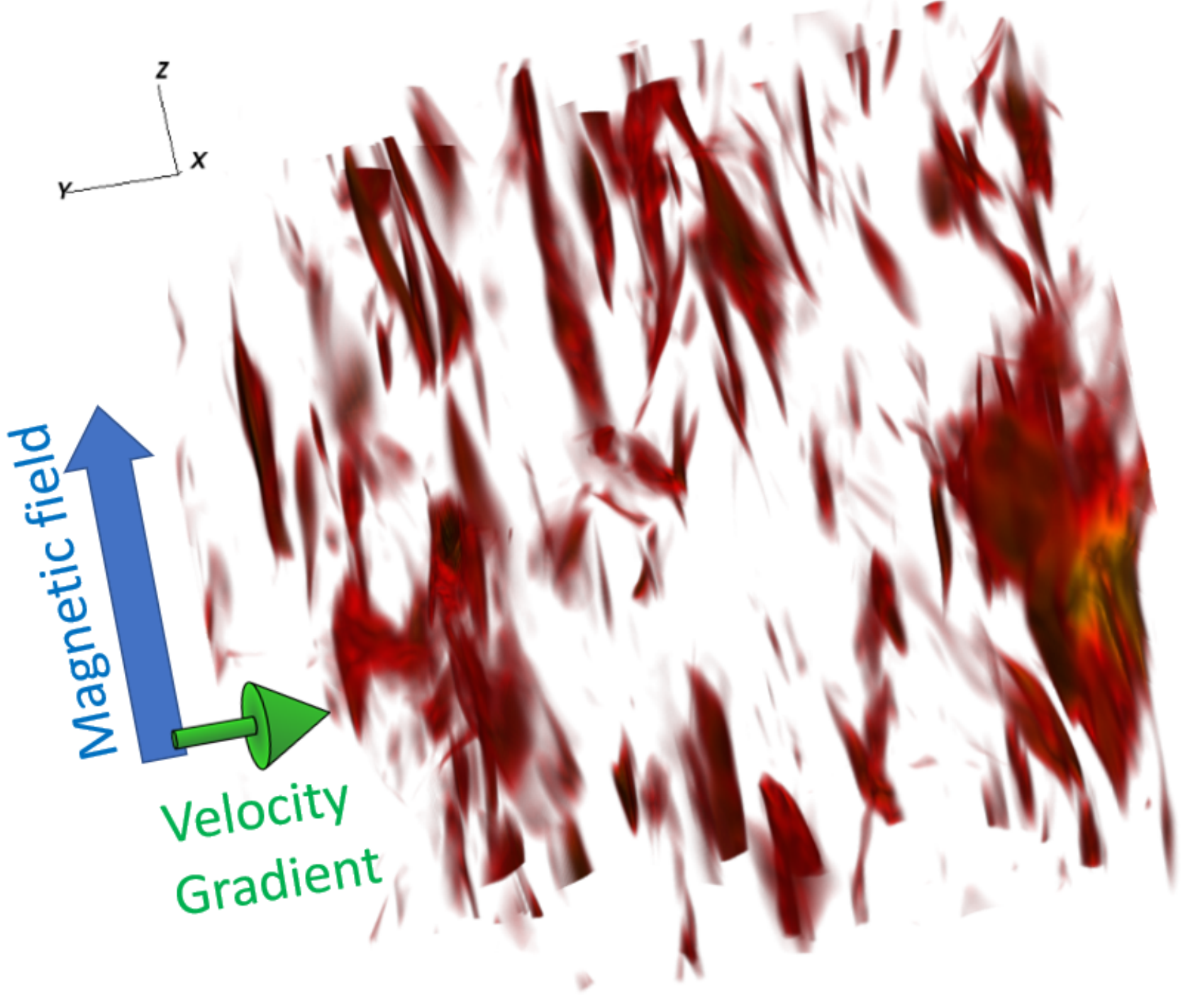}
\caption{\label{fig_illu} Iso-contours of equal velocities for sub-Alfvenic MHD turbulence simulation from our model Ma0.4 {\toreferee(See Table \ref{tab:sim})}. It is clear that gradients of velocity structures are perpendicular to the magnetic fields. From \cite{LY18a}.}
\end{figure}

{\toreferee In both sub-Alfvenic and super-Alfvenic turbulence the gradient methods of tracing magnetic field can be used (See \citealt{YL17b}). However, the signature of gradients in sub-Alfvenic and super-Alfvenic turbulence are different.} For {\toreferee sub-Alfvenic} turbulence, i.e. for $V_L<V_A$ the magnetic fields are always important and their action is imprinted on turbulent velocities at all scales. Figure \ref{fig_illu} illustrates the alignment of iso-contours of equal velocity (Red) and the mean magnetic field direction (Blue) in numerical simulations with $M_A=0.4$. {\toreferee (See \citealt{Chen16, Hull17})} It is obvious from the Figure \ref{fig_illu} that the direction of the fastest spacial change of the velocity is perpendicular to the magnetic field.

{\toreferee In the case of super-Alfvenic turbulence, extra complications in filtering large-scale isotropic eddies are required before applying the VGT method. } The cases of energy injection with $V_L<V_A$ and $V_L>V_A$  are discussed in Appendix \ref{App_A}. As we discuss there for the super-Alfvenic turbulence, i.e. for $V_L>V_A$ there is a range of scales for which turbulence is hydrodynamic and not much affected by magnetic fields. However, for turbulence with an extended inertial range there is a scale $L_{inj} M_A^{-3}$ at which the turbulent velocities get equal to the Alfven velocity (see Lazarian 2006). Starting from that scale our considerations above that relate the direction of velocity gradients and local magnetic field are applicable.

{\toreferee {\torefereetwo In the case of super-Alfvenic turbulence scales larger than $L_{inj} M_A^{-3}$, turbulence eddies are isotropic (See Appendix \ref{App_A}). As a result, one cannot derive velocity gradient information on these large scale eddies and therefore it is incapable to use the gradient method on isotropic large scale eddies to trace magnetic fields. }Turbulent eddies with scales smaller than $L_{inj} M_A^{-3}$ are still anisotropic and their gradients  still probes the local direction of magnetic field (See \citealt{YL17b}) . As a result, in the case of super-Alfvenic turbulence the gradients of velocities at large scale are not sensitive to magnetic field. However, this contribution can be spatially filtered out as it was demonstrated in LY17.}

\subsection{3D gradients induced by MHD turbulence}

With this understanding of turbulence in hand, it is easy to understand how the velocity gradient technique works. It is clear from Eq. (\ref{anis}) that the anisotropy of eddies increases with the decrease of the scale $l$. Thus the turbulent motions of eddies are getting better and better aligned with the local magnetic field as the scale of eddies decreases. This property of eddies is the corner stone of the velocity gradient technique that we introduced in a series of recent papers. 

In terms of practical measurements, it is easy to see that the gradients in turbulent flow increase with the decrease of the scale. This increase can be estimated as $v_l/l\sim l^{-2/3}$, where the Eq. (\ref{vel_strong}) is used. This suggests that the gradients arising from the smallest resolved eddies $l_{min}$ dominate the gradient measurements. When such eddies are well aligned with the direction of magnetic field, just turning the velocity gradients 90 degrees, one can trace magnetic field at the $l_{min}$ scale.

\subsection{Velocity gradients available through observations}

The 3D fluctuations of velocity are not directly available from observations. Instead, we demonstrated that gradients of velocity centroids \citep{GCL17,YL17a} and gradients of intensity fluctuations measured within thin channel maps \citep{LY18a} can be used as the proxies of velocity gradients. In both cases,  
the gradients are measured for a turbulent volume extended by ${\cal L} >L_{inj}$ along the line of sight and this entails additional complications, where ${\cal L}$ is the line-of-sight depth. While eddies stay aligned with respect to the local magnetic field, the direction of the local magnetic field is expected to change along the line of sight. Thus the contributions of velocity gradients gets summed up along the line of sight.\footnote{This effect is similar to that of the summation of far infrared polarization from aligned grains along the line of sight (see \citealt{Aetal15}) with the exception that the polarization {\toreferee sums quadratically (e.g. $Q\propto \int dz b_y^2-b_x^2$)in the form of Stokes parameters} while gradients are summed up as {\toreferee linear vectors (e.g. $\nabla C \propto \int dz \nabla (\rho v)$)}. The difference between the two ways of summation is small for small $M_A$, but gets significant for large $M_A$. Thus for super-Alfvenic turbulence we expect gradients to represent averaged along the line of sight magnetic field better compared to dust polarization.} 

\begin{figure}[t]
\centering
\includegraphics[width=0.49\textwidth]{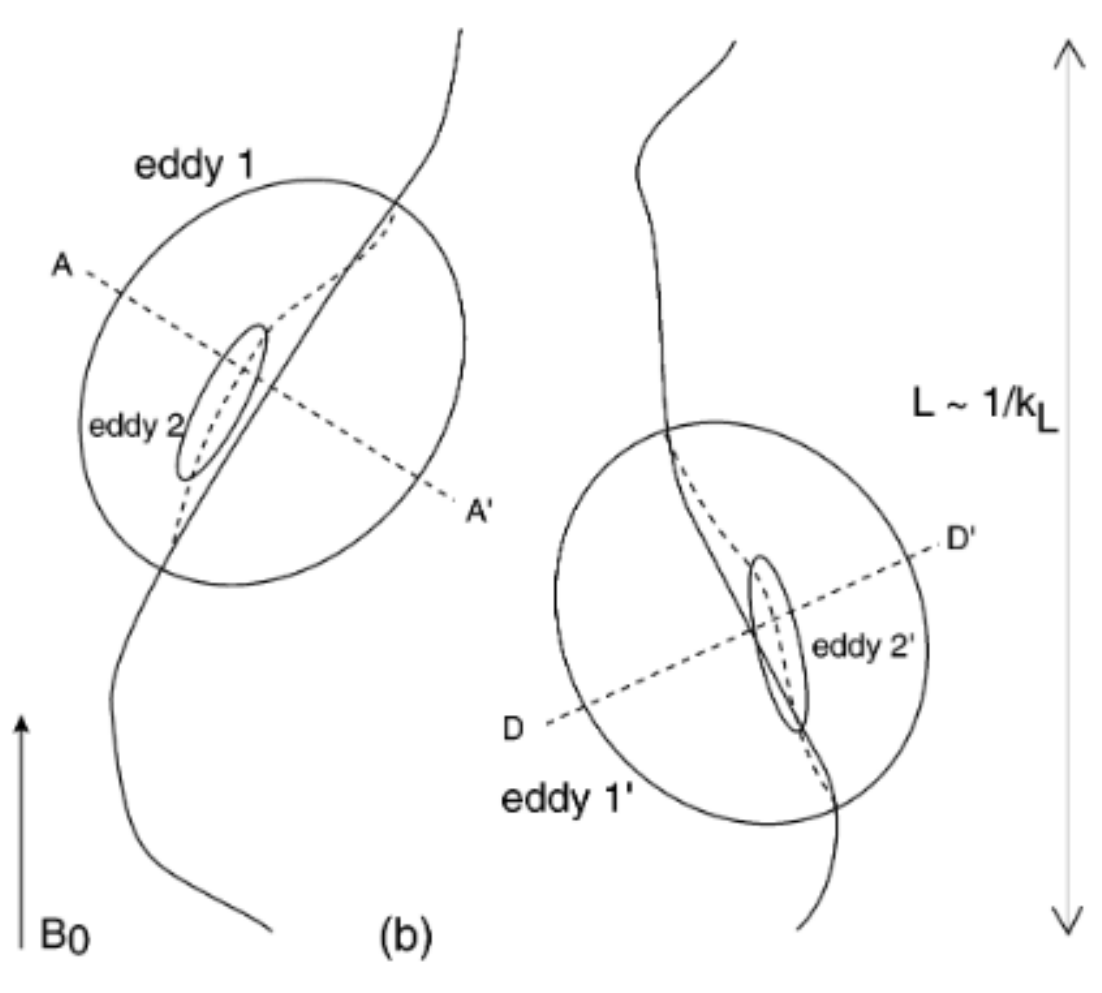}
\caption{\label{fig0} A pictorial illustration of the structure of anisotropic eddies. Small eddy 2 is aligned in respect to the magnetic field of the larger eddy 1. The same is true for the eddies 1' being aligned in magnetic field of eddy 2'. However, if measurements {\toreferee of eddies 2 and 2'} are made in respect to the mean magnetic field {\toreferee $B_0$,both eddies 2 and 2' will be treated as "misaligned" and contributes anonymously to the statistical tools (e.g. Correlation Function) with additional perpendicular-to-field contributions.} From \cite{CLV02}. }
\end{figure}

It is possible to show (see \citealt{LP12}) that for ${\cal L} >L_{inj}$ the local system of reference\footnote{\toreferee The importance of referring to the local system of reference is because it is the local magnetic field direction defines the direction of eddy anisotropy in magnetized turbulent media (LV99, Cho \& Vishniac 2000, Maron \& Goldreich 2001, Cho et al. 2002).} is not available from observations. Thus one has to use the reference system related to the mean magnetic field. In this system of reference, the anisotropy of the eddies is determined by the anisotropy of the largest eddies as it is illustrated in Fig. (\ref{fig0}). Indeed, the smallest eddies are aligned with the magnetic field and this magnetic field varies along the line of sight due to the the large scale variations of the magnetic field arising from the largest eddies (see \citealt{CLV02}). The latter variations are determined by the fluctuations {\toreferee of} the magnetic field variations at the injection scale, i.e. the changes in the direction of magnetic field along the line of sight are $\delta \varphi \approx \delta B/B\approx M_A$. We would like to stress that the tensors describing the fluctuations in local and global system of reference are different (e.g. compare the tensors in \citealt{CLV02} and  \citealt{LP12}). 

We have shown earlier that the 3D gradients of the value of 3D velocities is dominated by the smallest scales. A similar conclusion can be obtained for 2D observables. Indeed, the spectrum of observed fluctuations changes due to the line of sight averaging. It is easy to show that the 2D spectrum of turbulence obtained by projecting the fluctuations from 3D has the same spectral index of $-11/3$ (see \citealt{LP00}). The relation between the spectral slope of correlation function and the slope of the turbulence power spectrum in 2D in this situation is $-11/3+2=-5/3$, where 2 is the dimensionality of the space. Therefore, the 2D velocity fluctuations arising from 3D Kolmogorov-type turbulence  scale as  $l_{2D}^{5/6}$, with gradients anisotropy  scaling as  $l_{2D}^{-1/6}$. This means the contribution of the smallest scales is dominant for the measured 2D gradients.

\subsection{MHD turbulence anisotropies and distributions of gradient orientations}

Alfvenic turbulence is anisotropic with its anisotropy given by the relations that we have described above.
The anisotropy of Alfvenic turbulence imprints the scaling of Alfven modes into the anisotropy of slow modes (GS95, \citealt{LG01, CL02}). Alfven and slow modes together carry most of the energy of the turbulent cascade. Fast modes, in many cases, are sub-dominant in terms of the energy cascade, although they play a very important role for a number of key astrophysical processes, e.g. cosmic ray scattering (see \citealt{YanL02, YanL03, BL07}). The fast modes stay "isotropic"\footnote{We put isotropic in quotations, as the tensor describing fast modes is different from the tensor of isotropic turbulence (see more explanation in \citealt{LP12}).} both for pressure-dominated case (i.e. $\beta{\toreferee=2M_A^2/M_s^2}>1$, GS95) as well as magnetically-dominated case (i.e. $\beta<1$, \citealt{CL02}). In what follows we focus on the Alfvenic anisotropies that are measured in the observer's frame (see LP12). 

It is important to keep in mind  the statistical nature of turbulence. The velocity fluctuations have a distribution of directions about the magnetic field direction. Thus the gradients of velocities also have a distribution of directions, the peak of which {\toreferee is perpendicular to } the  magnetic field direction. The properties of the distribution are, however, very informative. It follows from the theory of Alfvenic turbulence that the dispersion of turbulence wave vector directions changes with the Alfven Mach number $M_A$ (see \citealt{LP16}). This should be reflected in the distribution of the directions of gradients and the goal of this paper is to determine this relation.  

Consider first the sub-Alfvenic and trans-Alfvenic turbulence.This degree of anisotropy is characterized by the ratio of the maximal to the minimal value of correlation function. This is because the relative orientation between the correlation direction and magnetic fields changes from parallel to perpendicular. Evidently for Alfven and slow modes, the correlation will be maximal along the direction of the magnetic field. It is clearly that there exists a correlation between the correlation function anisotropies and the gradients \citep{YLL18}. We shall also address the correlation in \S \ref{sec:result}.
\begin{table}[t]
\centering
\label{tab:sim}
\begin{tabular}{c c c c c}
Model & $M_S$ & $M_A$ & $\beta=2M_A^2/M_S^2$ & Resolution \\ \hline \hline
Ma0.2 & 7.31 & 0.22 & 0.01 & $792^3$ \\  
Ma0.4 & 6.1 & 0.42 & 0.01 & $792^3$ \\
Ma0.6 & 6.47 & 0.61 & 0.02 & $792^3$ \\  
Ma0.8 & 6.14 & 0.82 & 0.04 & $792^3$ \\  
Ma1.0 & 6.03 & 1.01 & 0.06 & $792^3$ \\
Ma1.1 & 6.08 & 1.19 & 0.08 & $792^3$ \\
Ma1.4 & 6.24 & 1.38 & 0.1 & $792^3$ \\
Ma1.5 & 5.94 & 1.55 & 0.14 & $792^3$ \\
Ma1.6 & 5.8 & 1.67 & 0.17 & $792^3$ \\
Ma1.7 & 5.55 & 1.71 & 0.19 & $792^3$ \\
 \hline \hline
\end{tabular}
\caption{Description of our MHD simulations.  $M_s$ and $M_A$ are the instantaneous values at each the snapshots are taken. This is refereed to Set C in the lower panel of Fig \ref{fig3}.}
\end{table}

{\toreferee From the discussion above, it is evident} that the maximum of the velocity gradients is perpendicular to the longest axis of eddies. This is true both for the 3D eddies and their line of sight projections. Within the procedure of block averaging \citep{YL17a}, the distribution of gradient orientation is fitted with a Gaussian function, where the peak of the Gaussian is associated with the direction of the velocity gradients within the block.

The Gaussian function used in \cite{YL17a} is a three-parameter function $A\exp(-\alpha(\theta-\theta_0)^2)$. Whether the other two fitting parameters are correlated to the intrinsic physical condition of MHD turbulence is unknown. However, it can be understood very easily that the height of the fitting function $A$ and the width $1/\sqrt{\alpha}$ are both related to the turbulence magnetization given by the Alfven Mach number $M_A$. Indeed, if the turbulent injection velocity $V_L$ is significantly small compared to the Alfven velocity, i.e. $M_A$ is small, then the bending of magnetic field is limited. The Alfvenic turbulence at small scales takes place through eddies that mix magnetic field lines perpendicular to the {\it local} magnetic field direction. For small $M_A$ all magnetic field lines are approximately in the same direction and therefore the dispersion of the velocity gradients that are perpendicular to magnetic field lines, is small. As the turbulence driving increases in amplitude the dispersion of velocity gradient orientations is expected to increase. In the framework of distribution of gradient orientations, a more prominent peak (i.e. higher $A$ and large $\alpha$) should be expected for small $M_A$.  Therefore one can infer the magnetization of a turbulent region by analyzing the amplitudes of $A$ and $\alpha$.

The magnetization can also be studied in the case of super-Alfvenic turbulence. As we discuss in Appendix \ref{App_A}, for scales larger than the scale at which turbulent velocity is equal to $V_A$  the correlation between magnetic field direction and the direction of the velocity fluctuations decreases. However, in this paper we are dealing with moderately super-Alfvenic turbulence and therefore the this correlation does not disappear completely. This allows to obtain $M_A$ from the distribution function of gradients. The latter is another name for distribution of gradient orientations that we will use further.   

\begin{figure*}[t]
\centering
\includegraphics[width=0.98\textwidth]{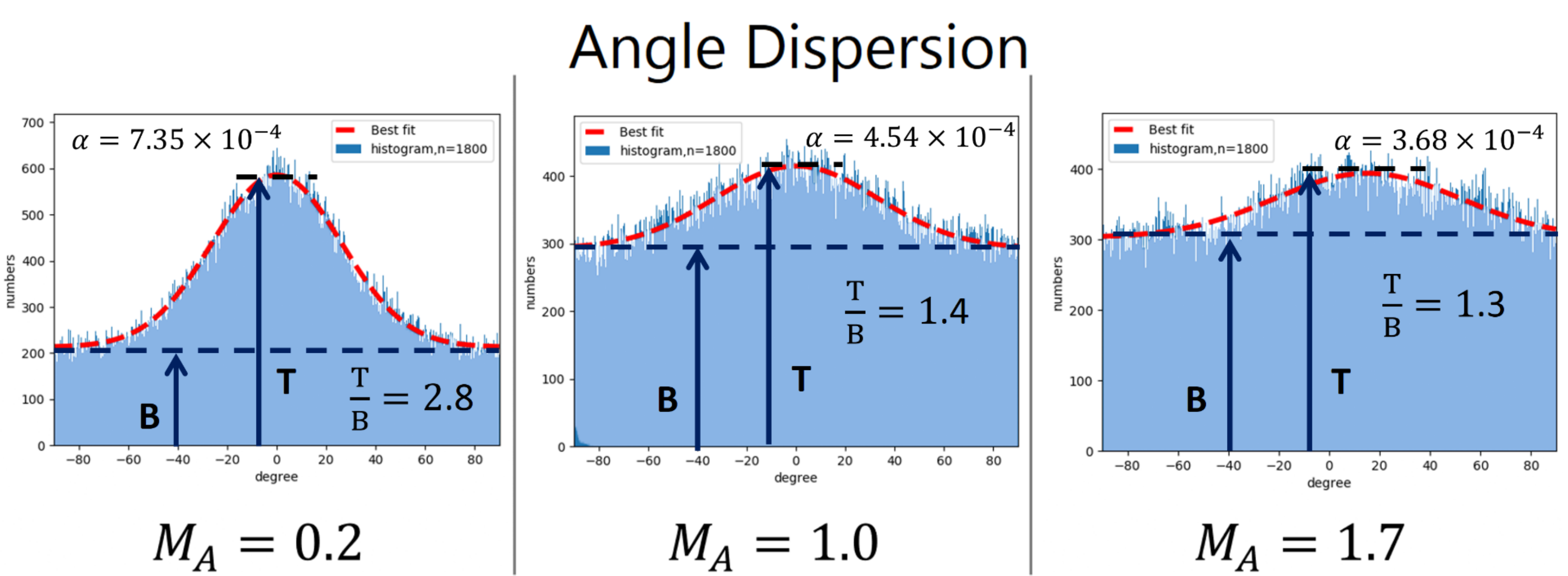}
\includegraphics[width=0.96\textwidth]{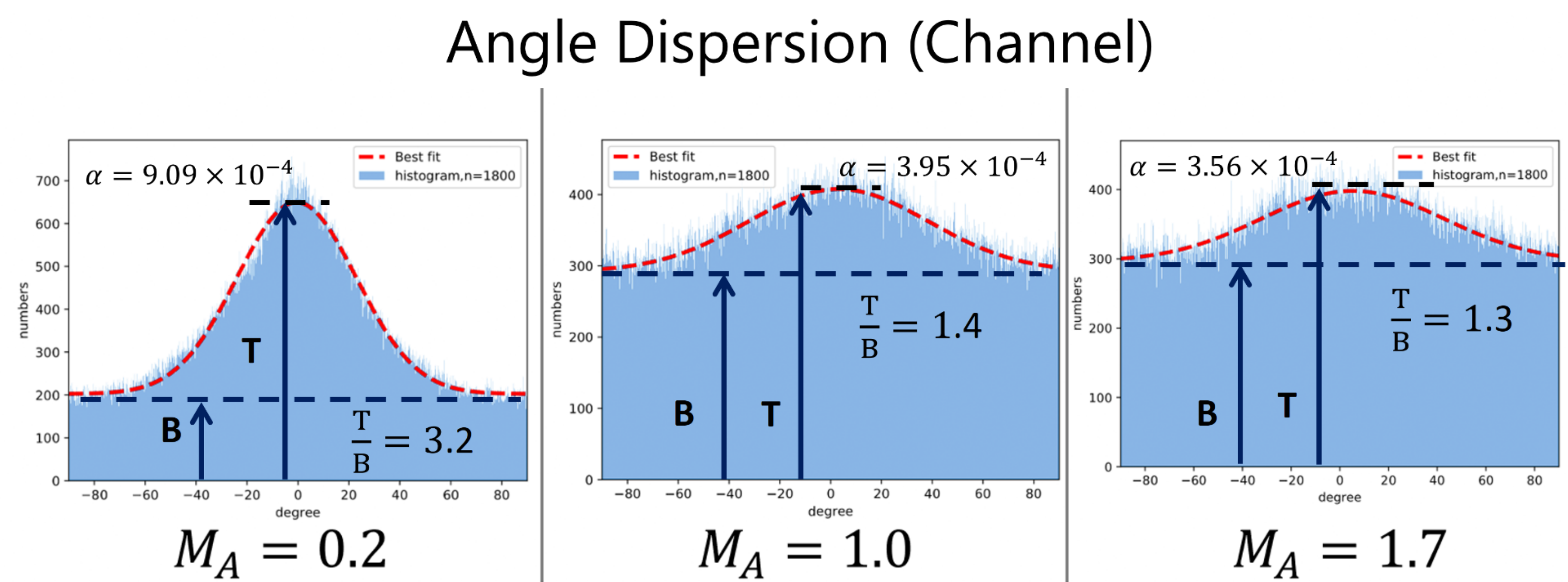}
\caption{\label{fig1} {\bf Upper Panel} The three panels here show the velocity centroid gradient (VCGs) orientation histograms from three numerical cubes with different Alfvenic Mach number $M_A$. We fit the three Gaussian with the modified Gaussian profile.{\toreferee  The T and B mean the peak and bottom value of the modified Gaussian profile which defined as  T = A + C'  and  B = C'.} {\bf Lower Panels} The same for the velocity channel gradients (VChGs). }
\end{figure*}

Fig. \ref{fig1} illustrates the distribution of gradient orientations for sub-Alfvenic, trans-Alfvenic and super-Alfvenic turbulence. The block averaging procedure \citep{YL17a} is used. The block size is chosen sufficiently large to allow Gaussian fitting. The latter is the necessary for the gradient technique that we have developed. The good Gaussian fit is important to know how reliable is the magnetic field direction that we determine with our technique and for determining the magnetization as we discuss in this paper. We clearly see that the width of the distribution is increasing with the increase of $M_A$. Similarly, the amplitude of the Gaussian distribution given by the ratio of the top-base values (see Fig. \ref{fig1}) is decreasing with $M_A$.  In the paper we are discussing the quantitative measures that can translate the parameters of the distribution of the gradient orientations into the $M_A$ data.

\subsection{Magnetic field and density fluctuations}

We believe that it is necessary to mention that our earlier studies have already shown that tracing of magnetic field directions is possible with magnetic gradients that are available, e.g. using synchrotron. The considerations about symmetry of velocity and magnetic fluctuations within Alfvenic turbulence constitute the basis for tracing magnetic field using not only with the Velocity Gradient Technique (VGT), but also the Synchrotron Intensity Gradients (SIGs) \cite{Letal17} and the Synchrotron Polarization Gradients (SPGs) \citep{LY18b}.

All our considerations above relevant to the distribution functions of velocity gradients are applicable to the SIGs and the SPGs. Therefore we expect to get  $M_A$ through studying the distribution of the gradient orientations of the synchrotron intensities and the synchrotron polarization. We defer presenting the corresponding results to other publications.  

Unlike the velocity and magnetic field fluctuations, the density fluctuations are not a direct tracer of MHD turbulence.  At small sonic Mach number $M_s$, the densities act as a passive scalar and follow the general pattern of velocity fluctuations \citep{2003MNRAS.345..325C,2007ApJ...658..423K}. The proxies available from observations are the Intensity Gradients (IGs). If those are  calculated using our recipes of block averaging for small $M_s$ they can also trace magnetic fields.  At the same time, being sensitive to shocks, the IGs can be successfully combined with the velocity or magnetic gradient measures, i.e. the {\toreferee Velocity Centroid Gradients (VCGs, \citealt{GCL17,YL17a}), Velocity Channel Gradients (VChGs, \citealt{LY18a}), Synchrotron Intensity Gradients (SIGs, \citealt{Letal17}) or Synchrotron Polarization Gradients (SPGs, \citealt{LY18b})}, for tracing both magnetic fields and shocks (see YL17b). 

As densities are sensitive to shocks we expect to observe that the distribution functions of  IGs to be both functions of $M_s$ and $M_A$, the dependence on the former being dominant for high $M_s$. We do not discuss the properties of the distribution functions of IGs in this paper.

\section{Numerical approach}
\label{sec:numerics}
We use 10 numerical cubes for the current study. Table \ref{tab:sim} presents the list of the MHD compressible simulations {\toreferee ,  some of which} were already used in \cite{YL17b,LY18a} . The latter {\toreferee paper provides } the details of the 3D MHD simulations employed.

For our set of simulations, the sonic Mach number is kept approximately constant ($M_s \sim 5.5 - 7.3$) {\toreferee since we are more interested in studying the effect of Alfvenic Mach number $M_A$ on the distribution functions of gradient orientations.} The chosen $M_s$ are {\toreferee within the range of sonic Mach numbers relevant to molecular clouds (Zuckerman \& Palmer, 1974)} .Observationally, one can approximate the $M_s$ by either studying the amplitudes of the velocity gradients (see \citealt{YLL18}) or investigating the width of the density probability distribution function \citep{BL12}.

Figure \ref{fig_illu} provides a visualization of velocity structures in one of our cubes. As we mentioned earlier, the iso-contours of equal velocity are aligned with magnetic field which makes the velocity gradients to be directed perpendicular to magnetic field. 

As we discussed earlier, the point-wise velocity information is not directly available. In this situation we calculate either velocity centroids or intensities within thin channel maps. Below we discuss both measures. 

{\bf Velocity Centroids}. The normalized velocity centroid $Ce({\bf R})$ in the simplest case\footnote{Higher order centroids are considered in \cite{YL17b} and they have $v^n$, e.g. with $n=2$, in the expression of the centroid. Such centroids may have additional advantages being more focused on studying velocity fluctuations.  However, for the sake of simplify we employ for the rest of the paper $n=1$.} and the emission intensity  $I({\bf R})$ are defined as
\begin{equation}
\begin{aligned}
Ce({\bf R}) &=I^{-1} \int \rho_v({\bf R},v) v  dv,\\
I({\bf R}) &= \int \rho_v({\bf R},v)  dv,
\end{aligned}
\label{centroid}
\end{equation}
where $\rho_v$ density of the emitters in the Position-Position-Velocity (PPV) space, $v$ is the velocity component along the line of sight and ${\bf R}$ is the 2D vector in the pictorial plane. The integration is assumed to be over the entire range of $v$.  The $Ce({\toreferee \bf R})$ is also an integral of the velocity and line of sight density, which follows from a simple transformation of variables (see \citealt{LE03}). For constant density this is just a velocity averaged over the line of sight. 

{\bf Fluctuations within Channel Maps}. Another measure that we employ is the channel map intensity. It is the emission intensity integrated over a range of velocities, i.e.
\begin{equation}
I_{\delta v}({\bf R}) = \int_{\delta v} \rho_v({\bf R},v)  dv,
\end{equation}
where $\delta v$ is the range of velocities for the integration. 

{\bf The Thin Channel Condition} If $\delta v$ is less than the velocity dispersion for the eddies under study, for such eddies the velocity slices were termed "thin" in \citeauthor{LP00} (\citeyear{LP00}, henceforth LP00, see also \citealt{LY18a}). For such slices the intensity fluctuations arising from the eddies are mostly induced by the velocity fluctuations (LP00). 

\section{Results}
\label{sec:result}
\subsection{ Determining $M_A$ with Distribution of gradient orientations}
\label{subsec:ma}
We investigate how the change of $M_A$ would alter the behavior of the distribution of gradient orientations. Distribution functions of the gradients for both centroids $Ce({\bf R})$ and velocity channel intensities $I_{\delta v} ({\bf R})$ are presented in Figure \ref{fig1}. {\toreferee These distribution functions are constructed by histograms with 1800 bins from three sets of numerical cubes with different $M_A$.} We fit the gradient orientation histogram using the Gaussian profile proposed in YL17a {\it by adding one constant shifting term}, i.e. 
\begin{equation}
F=A\exp(-\alpha(\theta-\theta_0)^2)+C'.
\label{F}
\end{equation}
 The value of the shift\footnote{In principle, the Gaussian profile with a shifting term is a better representation on the gradient orientation distribution in numerical simulations, since the {\toreferee histogram bin away from the peak} of gradient orientation distribution is usually much higher than zero. For instance, with $M_A\sim 0.2$ the velocity iso-contour axis-ratio can be in hundreds (Xu \& Lazarian 2018). However, simulations nowadays have limited resolutions. There is a natural tendency for velocity contours to have smaller axes ratio due to unresolvable minor axis. As a result, the {\toreferee histogram bin away} from the peak of gradient orientation distribution would not be close to zero. A constant shift would address the issue of finite length. In practice, the shift {\it will not} change the prediction of peak location by block-averaging.} $C'$and of the coefficient $A$  change  with $M_A$ which provides a way to study Alfven Mach number using the top-base ratio $(C'+A)/C'$. The fitting lines for each panel are shown in red dash lines in Fig \ref{fig1}. One can observe that the width of Gaussian profile increases with respect to $M_A$, while the top-base ratio decreases with the increase of $M_A$. The calculations were performed using block averaging (see \citealt{YL17a}) choosing blocks large enough to make sure that the Gaussian fitting is sufficiently accurate. Performing similar calculations for different simulations from Table 1 we observe noticeable changes in the distribution of gradients.  

The standard deviation of gradient orientation $\sigma_{GD}$ can be characterized by the circular statistics:
\begin{equation}
\begin{aligned}
\label{eq:gd}
\sigma_{GD} &= \sqrt{\log{\frac{1}{R^2}}}\\
&= \sqrt{-\log(\langle \cos \theta \rangle ^2+ \langle \sin \theta \rangle^2)}
\end{aligned}
\end{equation}
where $\theta$ is the gradient orientation of each pixel. Notice that in the formula, the range of $\theta \in [-\pi, \pi]$; but the observational gradient orientation in observational data (denoted as $\tilde{\theta}$) is in the range $[-\pi/2, \pi/2]$. In order to calculate the {\it inverted variance} $R=\sqrt{\langle \cos \theta \rangle ^2+ \langle \sin \theta \rangle^2}$, we  perform $\theta = 2\tilde{\theta}$ when processing the observational data. The quantity $1-R$ is called the {\bf variance} in circular statistics, which provides an alternative measure of dispersion for a set of directional data, and we shall use this parameter in the following sections. 

We would also evaluate the top-base ratio by dividing the peak value of the Gaussian peak from the base value in the gradient orientation histogram (See Fig. \ref{fig1}). We expect both of them would be related to the Alfvenic Mach number $M_A$ as we explained in \S \ref{sec:theory}. Fig. {\ref{fig1} illustrates pictorially how the fitting parameters for the distribution and the top-base ratio are obtained. Apparently, for low $M_A$ the distribution is strongly peaked and $A$ is large. This corresponds to good alignment of individual gradient vectors and the magnetic field direction. As $M_A$ increases, the gradient distribution is aligned with the magnetic field only in the statistical sense. This is especially obvious for $M_A>1$ as the velocity motions at large scales get uncorrelated with magnetic field. 

The panels in Fig \ref{fig2} compare the dependences of the variances and top-base ratios obtained for both centroid and channel gradients.  We observe that both these measures decrease as $M_A$ increases{\toreferee, which suggests that both top-to-base ratio and inverted variance are sensitive measures of $M_A$ in the range of $[0.2,1.7]$. }

In general, the findings for gradients in Fig \ref{fig1} and \ref{fig2} are easy to understand. They can be explained by a simple physical argument: When the magnetization is stronger (i.e. $M_A$ smaller)  , gradients tend to be more aligned with each other, therefore the peak of the histogram is more prominent (higher top-base ratio) and narrow (lower $\sigma_{GD}$). 

\begin{figure*}[t]
\centering
\includegraphics[width=0.96\textwidth]{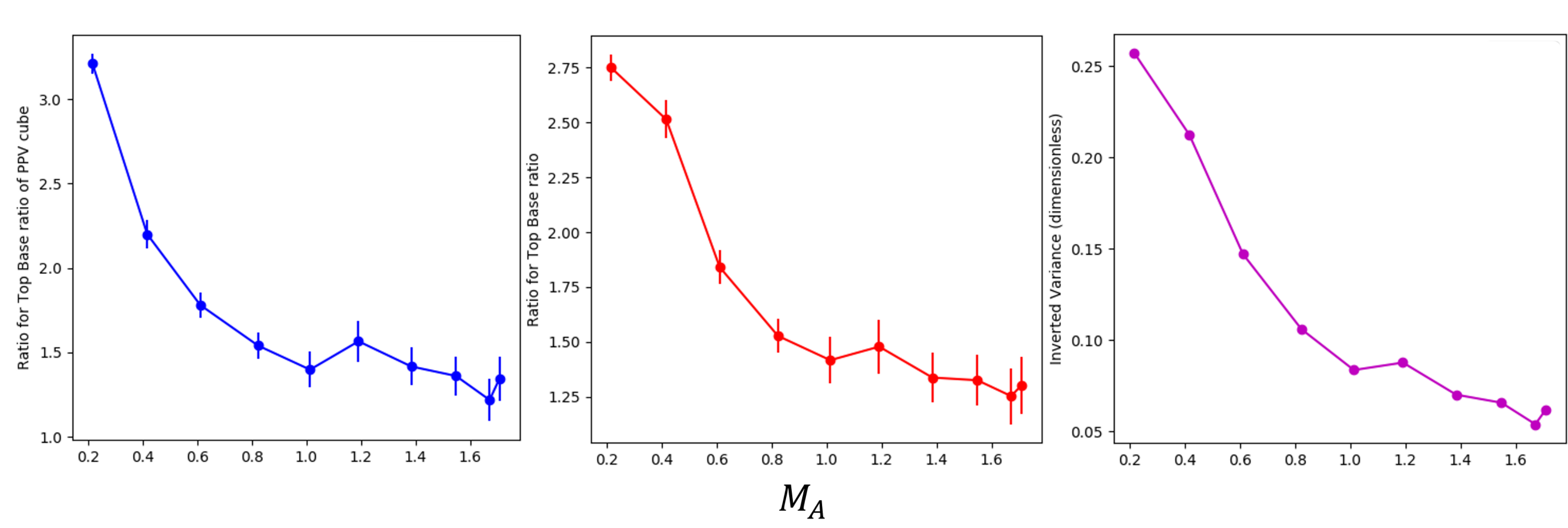}
\caption{\label{fig2} A comparison between the ways of measuring magnetization. From left to right:  Top-base ratio method for channels (Left) and Centroid (Middle), and the Inverted Variance $R$ method for Centroid (Right). The three quantities are plotted against $M_A$. There is a similar decreasing trend for the three method, especially for $M_A<1$. }
\end{figure*}

To extract the power-law dependency of $1-R$ to $M_A$, we plot the $(1-R)$ {\toreferee vs} $M_A$  relation (see left panel of Fig \ref{fig3}).  While there is a well-fit power law of $(1-R) \propto M_A^{0.14\pm0.03}$ , the fit is significantly flattened when $M_A \geq 1$.  
Similarly, the top-base ratio also showed a two-segment behavior in Fig \ref{fig3}. The fit of the top-base ratio to $M_A<1$ shows a power law ratio of $\propto M_A^{-0.46\pm 0.18}$.  The change of the power law index for $M_A>1$ is expected, as we have discussed earlier (see Appendix \ref{App_A}), the nature of turbulence is changing if the injection velocity gets larger than the Alfven speed. In this situation the large scale motions are dominated by hydro-type turbulence and the direction of magnetic field within the flow are significantly randomized. This changes the distribution function of gradient orientations. 

\subsection{Comparison with the polarization percentage technique}
\label{subsec:4.2}

\begin{figure*}[tbhp]
\centering
\includegraphics[width=0.98\textwidth]{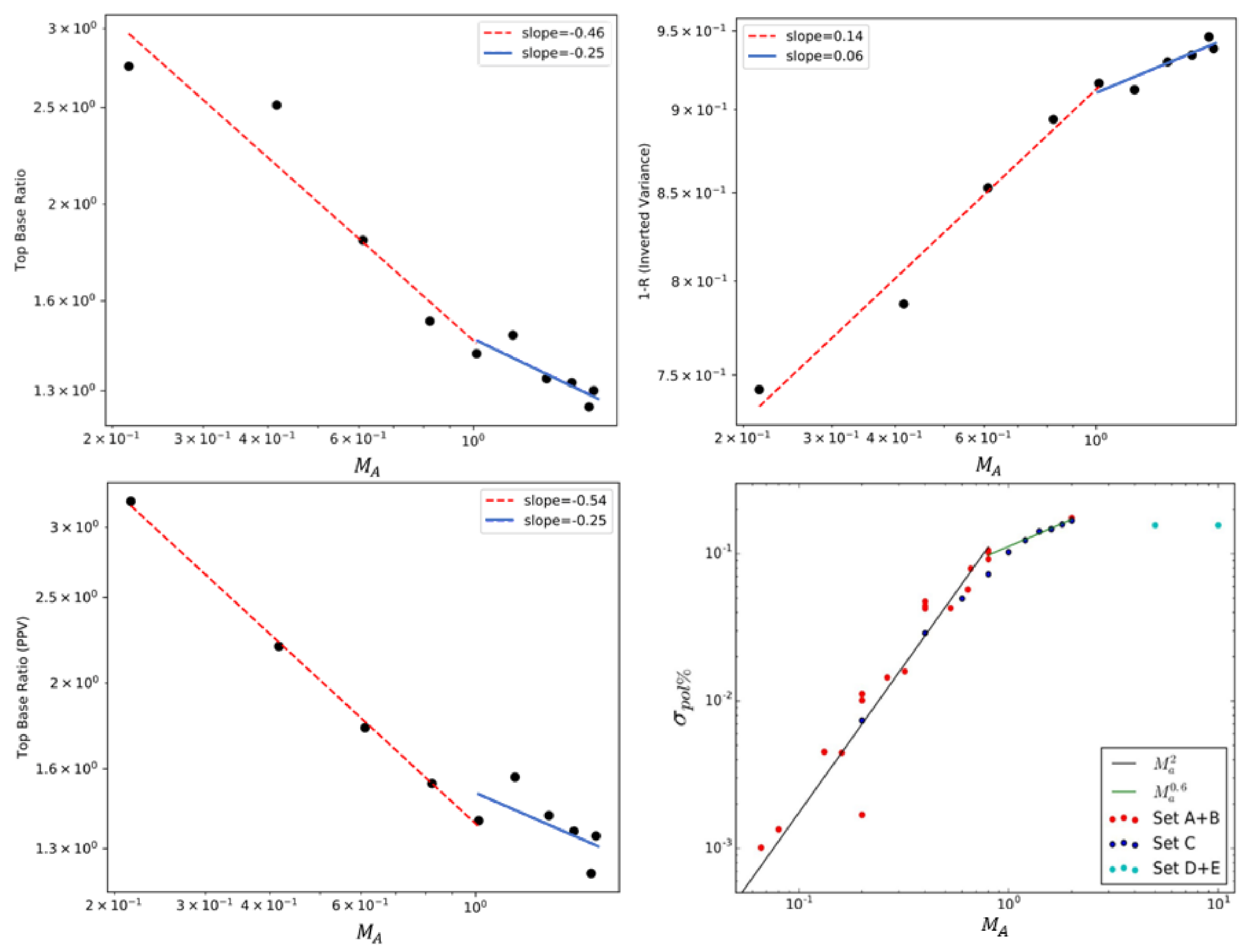}
\caption{\label{fig3}  (Top Left) The top-base ratio plotted against $M_A$ in log-log space {\toreferee using the simulation set from Table \ref{tab:sim}.} (Top Right) A log-log scatter plot {\toreferee using the simulation set from Table \ref{tab:sim}} showing the variation of $1-R$ to $M_A$. We fit the scatter plot with two segments, bending at $M_A\sim 1$. (Lower Left) A figure similar to the top-left corner but the data points {\toreferee from Table \ref{tab:sim}} computed from velocity channels instead of velocity centroids. (Lower Right)  A similar scatter plot with extra simulations in LY18 showing the general trend of $\sigma_{pol\%}$ to $M_A$ (See {\toreferee Appendix} \ref{app} {\toreferee \& \S \ref{subsec:4.2}.)} }
\end{figure*}

To illustrate the power of the techniques we showed in \S \ref{subsec:ma}, we would like to compare the results in  \S \ref{subsec:ma} to methods applicable to polarimetry. In {\toreferee Appendix} \ref{app} we provide a new way of evaluating $M_A$ using the dispersion of polarization percentage   $\sigma_{pol\%} $. While the traditional Chandrasekhar-Fermi technique uses the variations of {\it polarization directions} to determine $M_A$, the dispersion of polarization percentage does not require knowledge of circular statistics. In {\toreferee Appendix} \ref{app} we show that the analytical expectation that $\sigma_{pol\%}\sim M_A^2$ when $M_A<1$, which agrees well with our numerical calculations shown in the lower right of Fig \ref{fig3} obtained from 30 set of independent numerical simulations. For $M_A >1$ the dependence also changes similarly to the methods we suggested in \S \ref{subsec:ma}. For instance, the flattered fitting line with $M_A\in [1,2]$ has a proportionality of $\sigma_{pol\%} \propto M_A^{0.6}$. In comparison, for gradient dispersion, the power law is  $(1-R)\propto M_A^{0.06}$.

\begin{table}[h]
 \centering
 \label{tab:simulationparameters2}

 \begin{tabular}{c c c c c}
  Set & Model & $M_s$ & $M_A$ & Resolution\\ \hline \hline
  A & Ms0.2Ma0.02 & 0.2 & 0.02 & $480^3$\\
  & Ms0.2Ma0.07 & 0.2 & 0.07 & $480^3$\\
  & Ms0.2Ma0.2 & 0.2 & 0.2 & $480^3$\\
  & Ms0.20Ma0.66 & 0.20 & 0.66 & $480^3$\\
  & Ms0.2Ma2.0 & 0.2 & 2.0 & $480^3$\\
  & Ms0.20Ma0.66 & 0.20 & 0.66 & $480^3$\\
  & Ms0.02Ma0.2 & 0.02 & 0.2 & $480^3$\\
  B & Ms0.4Ma0.04 & 0.41 & 0.04 & $480^3$ \\
  & Ms0.8Ma0.08 & 0.92 & 0.09 & $480^3$ \\
  & Ms1.6Ma0.16 & 1.95 & 0.18& $480^3$ \\
  & Ms3.2Ma0.32 & 3.88 & 0.35  & $480^3$ \\
  & Ms6.4Ma0.64 & 7.14 & 0.66  & $480^3$ \\ 
  & Ms0.4Ma0.132 & 0.47 & 0.15  & $480^3$ \\
  & Ms0.8Ma0.264 & 0.98 & 0.32  & $480^3$ \\
  & Ms1.6Ma0.528 & 1.92 & 0.59& $480^3$ \\ 
  & Ms0.4Ma0.4 & 0.48 & 0.48& $480^3$ \\
  & Ms0.8Ma0.8 & 0.93 & 0.94 & $480^3$ \\ 
  & Ms0.132Ma0.4 & 0.16 & 0.49 & $480^3$ \\
  & Ms0.264Ma0.8 & 0.34 & 1.11  & $480^3$ \\ 
  & Ms0.04Ma0.4 & 0.05 & 0.52 & $480^3$ \\
  & Ms0.08Ma0.8 & 0.10 & 1.08 & $480^3$ \\ 
  D & d0 & 5.0 & 10.0  & $480^3$\\
   & d1 & 0.1 & 10.0 & $480^3,640^3,1200^3$\\
  E & e1 & 0.5 & 5.0 & $480^3$\\\hline \hline
\end{tabular}
 \caption {Extra simulation used for production of the lower right panel of Fig. \ref{fig3}. The Set C referred in Fig. \ref{fig3} is exactly what we showed in Table \ref{tab:sim}. }
\end{table}

\subsection{Comparison with other techniques for obtaining magnetization}

A number of other techniques for studying magnetization have been suggested. For instance, Tsallis statistics (see \citealt{EL10}) was shown to be sensitive to the value of $M_A$. In addition, one can estimate for sub-Alfvenic turbulence that $M_A\sim \delta \phi$, where $\phi$ is the dispersion of magnetic field directions. These directions can be obtained either through polarimetry studies, e.g. dust polarimetry, or using the variation of the velocity or magnetic field gradient orientations.\footnote{Variations of the synchrotron intensity gradients and synchrotron polarization gradients can also be used. The advantage of the gradient techniques is that they are independent of Faraday rotation. } 

{\toreferee To obtain $M_A$ through the Gaussian profile fitting requires that the conditions for the sub-block averaging in \cite{YL17a} are satisfied. In other words, the distribution of gradient orientation has to be well fitted by a Gaussian. Therefore the consistency of both directions (as traced by \cite{YL17a}) and magnetization from this work will be doubled checked through the sub-block averaging algorithm, i.e. whenever the Gaussian profile is not properly fitted there should not be any probe by VGT on neither magnetic field directions and magnetizations. While the velocity gradients present an independent technique for magnetic field tracing, it is synergetic with polarimetry measurements while dealing with the complex structure of interstellar magnetic fields.}


A more elaborated tool for studying $M_A$ is the analysis of the Correlation Function Anisotropy (CFA). This technique  has been explored in a number of publications \citep{EL11,BL12,Betal14,Eetal15,KLP16,KLP17a}, and we illustrate its results in the Appendix \ref{subsec:cfa}.

The advantage of the CFA technique is the existence of the analytical description that relates the measurements not only to $M_A$ but to the properties of the fundamental MHD modes, i.e. Alfven, slow and fast modes \citep{LP12,KLP16,KLP17a,KLP17b} However, the CFA technique requires calculating correlation functions and therefore is less local compared to the technique in this paper. Further research should reveal the synergy between different techniques of determining $M_A$. 

\subsection{Effects of adding noise}
\label{subsec:noise}
In this section we will show that the new technique of finding $M_A$ works in the presence of noise.  We add white noises with amplitudes relative to the standard deviation of Velocity Centroid $\sigma_C$ and test how the power law is changed as a function of noise amplitudes. In our test we apply white noise with amplitudes $0.1 \sigma_C$ and $0.2 \sigma_C$ and compare our results with the original fit we have in Fig \ref{fig3}. For noise suppression, we employ the Gaussian smoothing of $\sigma=2$ pixel as proposed in \cite{Letal17}. According to \cite{Letal17}, the kernel size we picked here would preserve the most small scale structures while efficiently suppress the noise in the synthetic map globally. By adding the noise and also the smoothing kernel, we can then test whether in noisy observations we can still use the top-base ratio and variance methods to estimate magnetization.

Fig \ref{fig4} shows the log-log  plot of both $(1-R)$ {\toreferee vs} $M_A$ and the top-base ratio to $M_A$ with noise addition (left) and smoothing (right). Before smoothing the noise-added $(1-R)$ {\toreferee vs} $M_A$ relation is very sensitive to the noise level, where only $0.2 \sigma_C$ can already flattened the plot. Fortunately, the application of the smoothing technique shows that the
magnetization tracing is not changed significantly. This gives us the confidence in applying our technique to observations. 

\begin{figure*}[tbhp]
\centering
\includegraphics[width=0.96\textwidth]{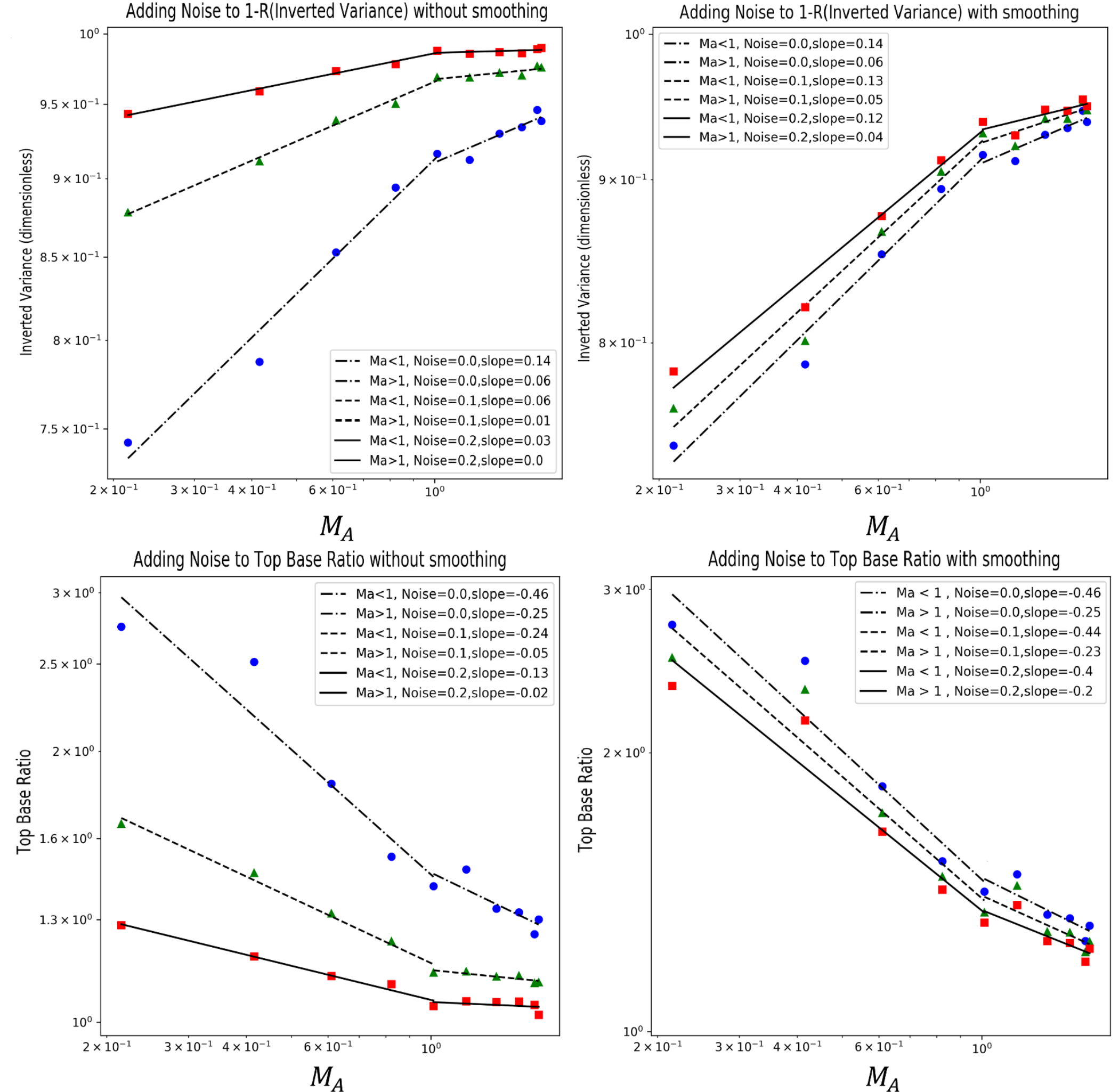}
\caption{\label{fig4} Four panels showing how noise and the suppression of it influences the result of magnetization tracing. For both methods of dispersion (top row) and top-base ratios (bottom row) we add white noise (Left column) to the map and see how the statistical parameters are altered. When the noise suppression techniques for gradients \cite{Letal17} are used, the trend of magnetization tracing (right column) is obviously more robust for both methods.}
\end{figure*}

\subsection{Observational data analysis}
\label{sec:observation}

We would like to demonstrate the application of the technique using GALFA-HI data. For details, please refer to the respective survey paper \citep{susantail}. The Galactic Arecibo L-band Feed Array HI (GALFA-HI) survey is an survey of the Galaxy in the $21 cm$ neutral hydrogen line. The data is obtained with the Arecibo Observatory 305 meter telescope. The telescope angular resolution  $~4'$. The region covers the sky at right ascension (R.A.) across $263.5^\circ - 196.6^\circ$, and declination (Dec.) across $22.5^\circ - 35.3^\circ$. Also, this piece of data covers a wide range of galactic latitude from middle to high, which is $26.08^\circ - 83.71^\circ$. {\toreferee The same region has been used in \cite{2015PhRvL.115x1302C} for the Rolling Hough Transform analysis.}

As we mentioned earlier, the choice of the block size is an important step for analyzing the data with gradients. If the block size is too small, the Gaussian fitting is poor which means that the determination of gradients is not reliable (see YL17a). If the block size is too big, the map is excessively coarse. Moreover, at large scales the regular variations of the direction of the Galactic magnetic field get important. As a result, the dispersion of angles increases with the block size. These considerations in \cite{LY18a} were used to optimize the block size\footnote{More sophisticated procedures have been also tested. For instance, one can filter out gradients with the largest and smallest amplitudes. The former corresponds to gradients arising from shocks (YL17b), the latter corresponds to noise. Fitting Gaussians into the remaining data can improve the quality of data. We plan to explore this and other ways of improving the gradient studies elsewhere. In this paper we use the block averaging as it is presented in YL17a.}  for the GALFA-HI map that we also use in this paper. 

We decompose the GALFA-HI map into blocks of size {\toreferee $150^2$ pixels, in which one pixel is 4',} and compute both the dispersion and top-base ratio of each block after a smoothing Gaussian filter of 4 pixels. Fig \ref{fig:susantail} shows the result of the $M_A$ distribution that we obtained with analyzing the distributions of gradient orientations. We clearly see similarity of the $M_A$ map {\toreferee obtained using } the top-base method and that obtained with the dispersion (in terms of the variance $1-R$) method. The fact that the two techniques provide very similar output, increases our confidence fin the distribution of $M_A$ that we obtain. 

In the top-base method, we marked the pixels as white (NaN), indicating that due to noise the gradient angle distribution could not be reliably fitted by a Gaussian profile. For such points the statistical variance was computed\footnote{\toreferee In principle, even the distribution of gradient orientations does not follow Gaussian, we can still compute the statistical circular dispersions (and thus inverted variance) according to Eq. \ref{eq:gd}. However whether the values computed through Eq. \ref{eq:gd} has any meaning in the framework of VGT is yet to be investigated.} using Eq. \ref{eq:gd}. This study is the first of this sort and has illustrative purposes. The accuracy of this approach for the regions where the Gaussian fitting is not working well will be studied elsewhere. {\toreferee For consistency to the Gaussian fitting requirement in \cite{YL17a}, we do not include the NaN data in the top-base method in the histogram in Fig \ref{fig:susantail2}.}

\begin{figure*}[tbhp]
\centering
\includegraphics[width=0.98\textwidth]{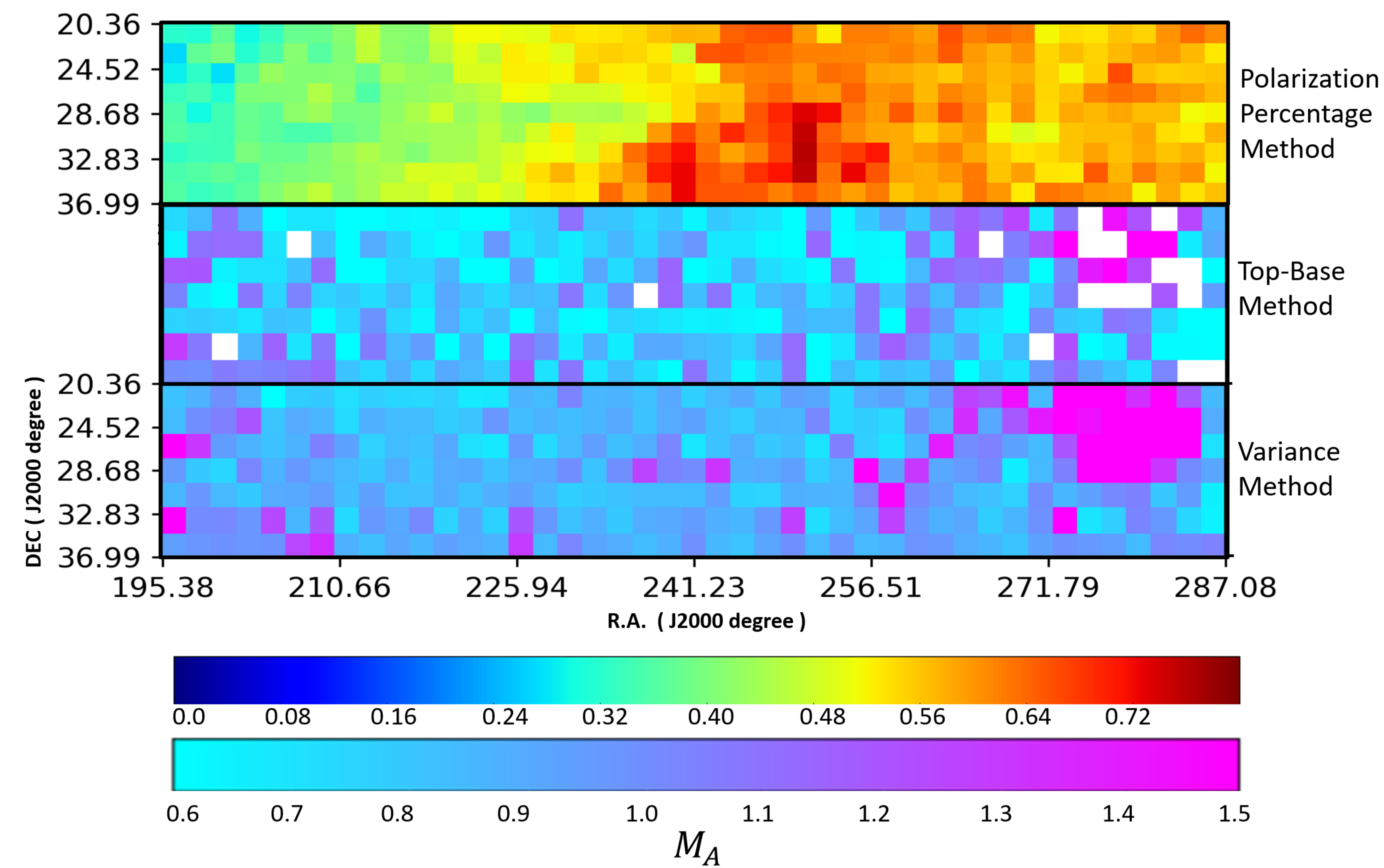}
\caption{\label{fig:susantail} 
 {\toreferee{\it Top Panel}}.The prediction of $M_A$ from the $P_\%$ method. {\torefereetwo As a matter of fact (e.g. \citealt{L07}),} the dust grain alignment drops significantly in low-latitude {\toreferee (approximately on the right-hand-side)} region, {\torefereetwo which explains why the expected $M_A$ in higher latitude will be significantly higher. Similar effect on polarization percentage is also observed in \cite{Planck1519}. }
 {\toreferee \it Middle Panel}. $M_A$ distribution obtained using top-base ratio approach from this paper on the region of GALFA-Hi with block size of $150^2$ pixels. The data was from \cite{2015PhRvL.115x1302C} and \cite{susantail}.
The white pixels are the regions that cannot be fitted by the Gaussian profile. {\toreferee \it Bottom Panel} The distribution of $M_A$ obtained using the method of variance. There is a similarity of the $M_A$ predictions between the {\torefereetwo top-base ratio and the variance method}.}
\end{figure*}

{\toreferee We also test} how the selection of block size changes our estimate of $M_A$.  Fig \ref{fig:susantail2} shows the normalized distributions of $M_A$ of the piece of GALFA region (Fig \ref{fig:susantail}) when we vary the block size of calculating the top-base ratio. While there is a fluctuation of magnetization estimation for different block size, both the peak value and the shape of the distributions are very similar. One can observe that the peak values of these distributions are all around $M_A\sim 0.75$, and exhibit a rough Gaussian profile. This suggests that the block size does not significantly change  the estimate of $M_A$. However, we expect that the regular curvature of magnetic field lines within the block{\toreferee, which induced a broaden distribution of gradient orientation,} is a factor that makes us {\toreferee difficult in estimating $M_A$} in high latitude HI (see also lower $M_A$ obtained with for the same data with the gradient amplitude method in \cite{YLL18}). The issue of obtaining the accurate values of $M_A$ {\toreferee in regions with regular curvatures} definitely requires further detailed studies.

\begin{figure*}[tbhp]
\centering
\includegraphics[width=0.98\textwidth]{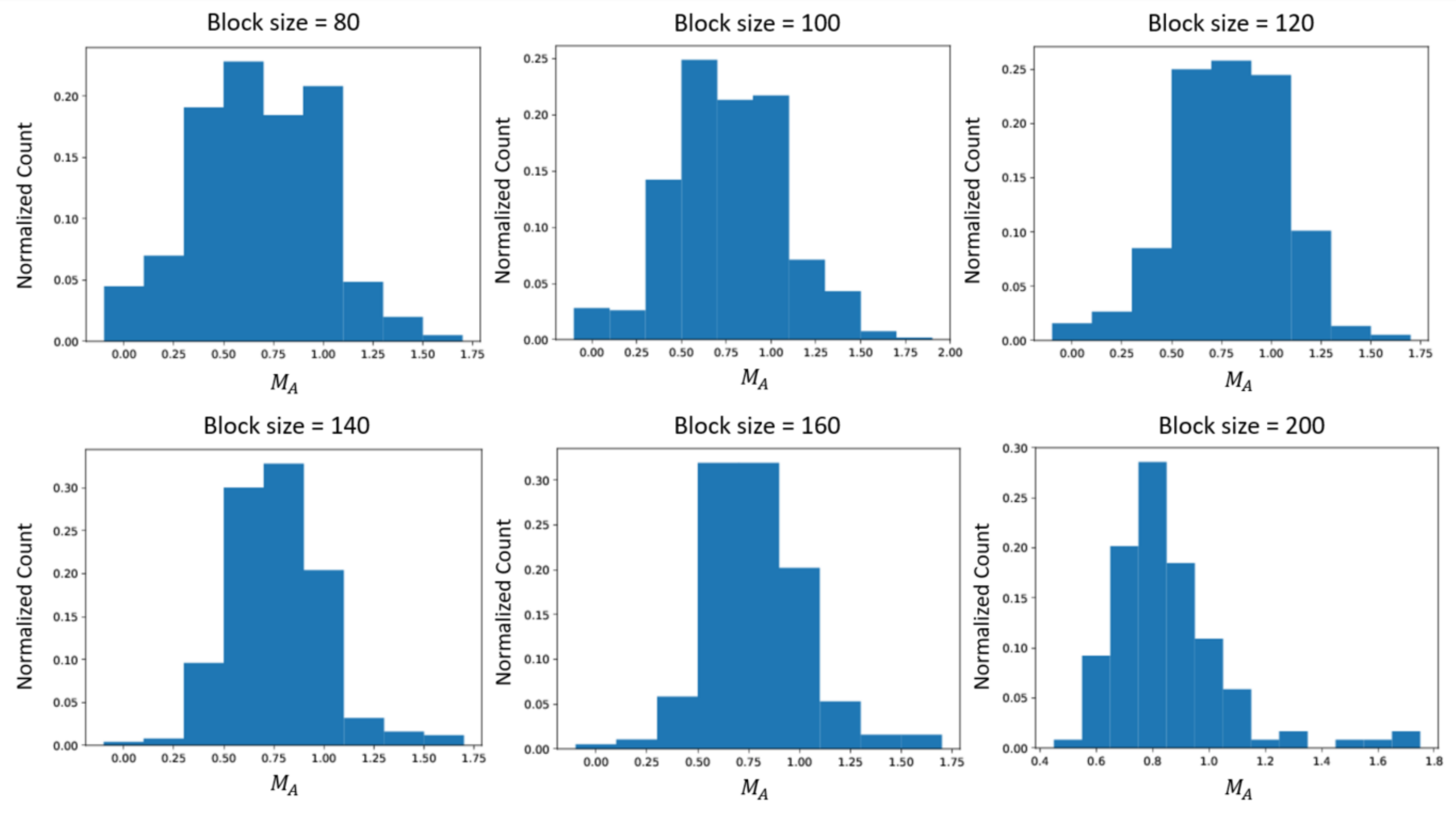}
\caption{\label{fig:susantail2} The estimated $M_A$ for Fig \ref{fig:susantail} using the top-base ratio method in different block sizes. The peak ($M_A\sim 0.75$) remains unchanged even through the sampling area increases for about $2.5^2\sim 6$ times.}
\end{figure*}

\section{Obtaining magnetic field intensity distribution from the $M_A$ distribution}
\label{intensity}

The Alfven Mach number $M_A$ is an essential characteristic of magnetized turbulent media that is important for describing key astrophysical processes, e.g. an {\toreferee explicit} discussion of the dependence of cosmic ray propagation in  \cite{LY14, L16} and heat conduction in \cite{L06}. 

{\toreferee In principle, if one can obtain accurate polarization measurements with sufficiently high dust grain alignment efficiency (See Lazarian 2007), then the method of gradients has no particular advantage in front of polarization measurements. However, in observational environments the polarization fraction is fairly low (e.g. in case of the polarization measurements near molecular cloud centers), which will often lead to insufficient statistics in computing the dispersion of polarization angles. Comparatively, the VGT is versatile in different physical regimes \citep{YL17b,LY18a} and thus the estimation of magnetic field informations (directions and more importantly magnetization) can be applied to a wider regime without any statistical problems we faced in polarization measurements.}

In addition, if the Alfven Mach number is known, it is possible to obtain magnetic field strength from the relation $M_A=\frac{\delta B}{B}$. Assuming that the relation of $\delta B$ and the dispersion of the velocity obeys the Alfvenic relation, i.e. $\delta V=\delta B/(4\pi \rho)^{1/2}$ one can express the magnetic field strength as 
\begin{equation}
B_{POS}=\sqrt{4\pi \rho} \frac{\delta V}{M_A}
\label{BB}
\end{equation}
where $\delta V$ can be associated with the velocity dispersion and we took into account that it is the plane of the sky (POS) component of magnetic field that is being explored with the technique. As only line of sight (LOS) velocity $\delta v_{los}$ is available, for the practical use of the Eq. (\ref{BB}) the velocity dispersion $\delta V$ there should be associated with $\delta v_{los}$, i.e.  $\delta V=C\delta v_{los}$ , where $C$ is a coefficient that relates the dispersions of the turbulence POS velocities with the available LOS ones. For an uniformly distributed Alfven wave that moves along the mean magnetic field line, $C=0.5$. The coefficient $C$ grows up when the angle between the mean magnetic field direction and the line of sight is smaller. In the limiting case of magnetic field parallel to the line of sight, $C$ is not defined, as no LOC velocities can be measured.

In reality, the observed magnetic field is not simply composed of a dominant mean field with straight-line morphology and infinitesimal Alfven waves moving along the mean field. For realistic turbulence the magnetic field wandering is significant (see LV99, \citealt{ELV11}). Moreover, MHD turbulence consists of three MHD cascade modes, namely the Alfven, slow and fast modes  which every mode has unique cascade properties (see \citealt{LG01,CL02}). As a result,  in most practical studies the coefficient $C$ is treated as an empirically given parameter.\footnote{Analytical studies in \cite{KLP16} open a way to calculate $C$ from the first principals for the given level of turbulence, inclination angle of the mean magnetic field and the line of sight and the assumed composition of fundamental MHD modes. Incidentally, an additional modification of Eq. (\ref{BB}) is required to account for {\toreferee the} other types of modes {\toreferee that are also present} in the magnetized turbulent flow.}

In addition, the calculations are a bit more complicated for the case of small scale turbulence injection (see \citealt{YC14}). We discuss the details of obtaining magnetic field intensity from observations elsewhere. For the time being, it is important that Eq. (\ref{BB}) provides a rough guidance for exploring magnetic field strength. 

\section{Discussion}
\label{sec:discussion}

\subsection{{\toreferee Broader} applications to other types of gradients}

The cornerstone of the present work is the Gaussian fitting function first used in \cite{YL17a}. This provided us with the ways of obtaining the dispersions and the top-based ratios that are used to trace $M_A$. While we did all the estimations of $M_A$ for velocity centroids {\toreferee in the present paper}, we expect the technique presented in the current work will be applicable to other measures, e.g, reduced centroid gradients from \cite{LY18a}, synchrotron intensity gradients from \cite{Letal17}, synchrotron polarization gradients and synchrotron derivative polarization gradients from \cite{LY18b}, and IGs from \cite{YL17b}. {\toreferee More importantly, we showed in Fig \ref{fig1} that in dispersion of velocity channel gradients also has the same trend as the centroid gradients, which signifies the dispersive quantities we introduced in the current work should also work on other gradient measures. }

If turbulent velocity broadening is known, then, as we discussed \S \ref{intensity}, the magnetic field strength can be obtained. However, by itself $M_A$ is a key parameter describing astrophysical turbulence and therefore its  determination of its distribution is essential for many astrophysical processes. 

\subsection{The improved techniques of gradient calculation and use of interferometers}

By now we have obtained a set of tools developed in the Velocity Gradient Technique \citep{GCL17,YL17a,YL17b,LY18a}, e.g. error estimations, wavenumber filtering, identifying shocks and gravitationally-bound objects etc. Some of these tools have been already applied to the magnetization tracing technique discussed in the present paper. In particular, the noise suppression technique proposed in \citep{Letal17} is shown to be effective in maintaining the established power-law relation in Fig \ref{fig4}. Some of these tools we apply as the technique matures. For instance, in \cite{LY18a} we demonstrated  how one can improve the tracing of magnetic field with velocity gradients. This approach based on the moving window technique, seems promising  for magnetization studies. 

To get higher resolution maps of magnetization it is advantageous to use interferometers. We have shown in our previous publications (e.g. \cite{Letal17}) that to trace magnetic field with gradients one can successfully use interferometers. It is important to understand that the interferometric data can be used directly and it is not required to have all spatial freuencies to use our gradient technique. Indeed, the gradients are calculated for smallest separations and, in fact, filtering of low spatial frequencies is recommended for increasing signal to noise ratio. Therefore it is not surprising that the structure of magnetic field can be reproduced e.g. without employing single dish observations. It is easy to see that the same statement is true for using the techniques presented in the present paper. This opens exciting perspective for of the mapping the distribution of $M_A$ for external galaxies.  

\subsection{Studying 3D magnetization}

Velocity gradients provide a way to probe turbulence and establish the distribution of magnetic field directions in 3D. For instance, this can be achieved by using the galactic rotation curve for atomic hydrogen of the Milky Way. For molecular clouds different transitions from multi-molecular species (e.g. 12CO, 13CO) can provide information of magnetic field structures at different depths {\toreferee suggesting} that not only the directions, but also 3D magnetization can be traced by velocity gradients {\toreferee through stacking the magnetic field maps from molecular tracers with different optical depth}.  Similar idea of exploring the 3D field tomography through velocity gradients with rotation curve is explored in \cite{CL18}. Together with the CF-method, the distribution of 3D magnetic field strength can also be acquired (see \S \ref{intensity}).

MHD turbulence imprints its properties on the statistics of synchrotron intensity/polarization gradients as well as on the gradients of emission intensity. In terms of utilizing magnetic fluctuations, the sister techniques are available. They are the Synchrotron Intensity Gradient (SIG) technique and  \citep{Letal17} and the Synchrotron Polarization Gradient (SPG) technique \citep{LY18b}. These techniques are demonstrated to trace the magnetic field {\toreferee in the plane of sky}. In particular, the synchrotron polarization gradients \citep{LY18b} can recover the 3D distribution of magnetic field directions by using the effect of Faraday depolarization. Our present results {\toreferee of tracing $M_A$} with the width and the top-base ratio for the gradient distribution are also applicable to the techniques using synchrotron. In terms of the SPG technique, this provides another way of studying 3D magnetization. Combining the results on magnetic directions and magnetization obtained with the gradients of synchrotron and spectral lines, one can get an unprecedented insight into the magnetic structure of the multi-phase ISM. 

Gradients of intensities of gas or dust emission\footnote{\toreferee By intensity gradients we understand the measures calculated using the block averaging procedure described in \cite{YL17a} and the resulting Intensity Gradient Technique (IGT) is different from the Histograms of Relative Orientation technique in Soler et al. (2013). A comparison between the two techniques is provided in our earlier publications on gradients as well as in our forthcoming paper containing a very detailed quantitative comparison of the IGT and the HRO.} provide additional information about the ISM.  For instance, the gradients of intensities are strongly affected by shocks (see \citealt{YL17b}). This opens interesting prospects of studying shocks and sonic Mach numbers by comparing the {\toreferee distribution of gradient orientations} of velocities and intensities. We shall explore this possibility elsewhere.

\subsection{Calculating the distribution of expected dust polarization}

Magnetic field tracing is routinely done with dust polarization. It is frequently assumed that the observed polarization represents projected magnetic field weighted by gas density. This is approximately true when two conditions are simultaneously satisfied: (1) A {\toreferee relatively high} grain alignment {\torefereetwo (greater than a few percent)}; (2) magnetic fields {\toreferee in the plane of sky} does not change significantly along the line of sight. While the former is mostly satisfied in the diffuse interstellar media (see \citealt{L07} for a review), the latter is a more subtle requirement. The problem arises from the fact that the additions of the polarization and magnetic field follow different laws, which is usually under-appreciated by observers. Polarization summation is a summation of quadrupole quantities, which is different from the summation of vectors of magnetic field. The difference become obvious when the direction of magnetic field lines are oscillating along the line of sight. 

Velocity gradients are linear vector quantities similar to magnetic fields. However, the addition of the gradients along the the line of sight is happening in the random walk fashion due to the absence of the {\toreferee{\it direction}(i.e. the vector-head)} of gradients {\toreferee by} the symmetry of the anisotropic eddies. The addition method of velocity gradients is not only different from that of magnetic field, but also for polarization. We feel that gradients {\toreferee can} present a better representation of the projected magnetic field for the case of super-Alfvenic turbulence, i.e. when magnetic field directions are changing strongly along the line of sight,  as both gradients and magnetic field tend to cancel {\toreferee themselves out respectively.} 

The issue of the comparison of the gradients versus polarization as a representation of magnetic field is an issue of our separate paper (Yuen \& Lazarian, in prep). Instead we will make remarks about the sub-Alfvenic case where the differences in the magnetic field direction along the line of sight is limited. In this case the variations in the measured degree of polarization reflect the variations magnetic field directions. These variations are also reflected by the dispersion of gradients that we deal in this paper. Therefore measuring the aforementioned dispersion one can predict the degree of polarization expected in the given direction. In other words, at least for sub-Alfvenic turbulence velocity gradients {\toreferee can} predict both the direction of polarized radiation and the degree of polarization. This finding is important for CMB polarization foreground studies where it is good to have an independent way of finding the expected polarization. 

\subsection{Comparison with other works}
 Our present study reveals a new valuable feature of gradients, i.e. their ability to deliver the value of magnetization through studies of the distribution of their directions within the block over which the averaging is performed. The theory of MHD turbulence predicts a distribution of wavevectors that depends on the ratio of the ratio of the turbulent to magnetic energies $\sim M_A^2$. Therefore it is only logical that we find that the distribution of the gradient orientations  also depends on $M_A$.
 
This provides another way of using the information obtained with the gradient technique in order to study properties of magnetized interstellar medium. Our earlier studies (e.g. \citealt{YL17a,LY18a}) provided ways of magnetic field tracing with the block-averaged velocity gradient orientations as well as of obtaining the sonic Mach number $M_s$ via studying the gradient amplitudes \citep{YLL18}. Therefore combined with the results of the present paper, the study of velocity gradients can provide both $M_A$, $M_S$ as well as magnetic field direction.\footnote{We would like to stress that similar results can be obtained with synchrotron intensity and synchrotron polarization gradients.  techniques. } These quantities can be studied not only in 2D, but also in 3D. This brings magnetic field studies to a new level.
 
 We would like to emphasize that what we suggest in the present paper should not be confused with the studies of Histograms of Relative Orientation (HRO) technique in \cite{Setal13} and subsequent works (e.g. \citealt{Planck35}). In this paper we discuss the distribution of orientation of velocity gradient orientations around the block-averaged \citep{YL17a} direction of the velocity gradient. The block averaged direction in our technique determines the magnetic field orientation. On the contrary, HRO (a) it is not capable to trace magnetic field direction, it relies on polarimetry to do this, (b) it uses density gradients, not the velocity gradients, (c)  it is not capable of revealing $M_A$ and $M_s$. The value of HRO is obtaining the statistical correlation of the averaged density gradient orientation and magnetic field as a function of the column density. All in all, HRO is a different technique introduced with a different purpose. As we mentioned earlier, our approaches of block averaging are applicable to density gradients and this is the basis of our intensity gradients technique (IGT). In terms of its comparison with the HRO it does not have the difference given by item (b), while in terms item (c),  our research shows that apart from being sensitive to $M_A$ and $M_s$ the IGT can identify shocks.  The use of the IGT and velocity gradients is synergistic as it helps to reveal the regions of gravitational collapse \citep{YL17b, LY18a}.

\section{Summary}
\label{sec:summary}

The present paper {\toreferee establishes} a new way to study the ISM magnetization. We characterize the magnetization by the Alfven Mach number $M_A$, the value of which is important for solving many astrophysical problems. To find $M_A$ we use the properties of the distribution of the gradient orientations, namely, the velocity gradient dispersion and the top-base ratio obtained through the Gaussian fitting of the distribution \citep{YL17a}. To summarize:
\begin{enumerate}
\item We establish the power-law relations between the statistical parameters of the distribution of gradient orientations, i.e. the variance $1-R$ and the Top-Base ratio,  to the Alfvenic Mach number $M_A$ (\S \ref{sec:result}). 
\item We discuss a possibility of using the galactic rotation curve and different spectral lines to get the 3D map of the $M_A$ distribution.
\item We show that combining $M_A$ with the  dispersion of the Doppler-broadened spectral line ,  one can acquire the magnetic field strength {\it without using the polarimetry}.
\item Our method is consistent with the method of correlation function anisotropy (CFA) in tracing $M_A$ (\S \ref{subsec:cfa}). We show that the gradient technique can provide maps of $M_A$ with higher resolution.
\item We show that our technique of  $M_A$ tracing is a robust tool in the  presence of noise (\S \ref{subsec:noise}).
\item We applied our technique to HI observational data (\S \ref{sec:observation}) to obtain the distribution of $M_A$ over an extensive region of the sky. 
\item Our approach for finding $M_A$ using the dispersion of gradient distribution is applicable not only to velocity gradients, but also to magnetic gradients that are measured with synchrotron intensity gradients and synchrotron polarization gradients.
\end{enumerate}

{\bf Acknowledgements.}   
We thank Susan Clark for providing us with the HI data and friendly discussions. This publication utilizes data from Galactic ALFA HI (GALFA HI) survey data set obtained with the Arecibo L-band Feed Array (ALFA) on the Arecibo 305m telescope. The Arecibo Observatory is operated by SRI International under a cooperative agreement with the National Science Foundation (AST-1100968), and in alliance with Ana G. Méndez-Universidad Metropolitana, and the Universities Space Research Association. The GALFA HI surveys have been funded by the NSF through grants to Columbia University, the University of Wisconsin, and the University of California. A.L. acknowledges the support from grant NSF DMS 1622353 and ACI 1713782. {\torefereetwo We thank the anonymous referee for providing extensive important comments and suggestions on our work.}

\appendix

\section{Gradients for Sub and Super Alfvenic turbulence}
\label{App_A}

In the main text we described the case of trans-Alfvenic turbulence  corresponding to the injection of energy at the scale $L_{inj}$ with $V_L=V_A$. If the energy is injected {\toreferee with the injection velocity $V_L$ that is} less than {\toreferee the} Alfven speed {\toreferee $V_A$}, the turbulence is {\it sub-Alfvenic}. In the opposite case it is {\it super-Alfvenic}. {\toreferee The illustration of turbulence scalings for different regimes can be found in Table\ref{tab:regimes}. We briefly describe the regimes below. A more extensive discussion can be found in the review by Brandenburg \& Lazarian (2013). 

\begin{table*}[t]
\caption{Regimes and ranges of MHD turbulence. \label{tab:regimes}}
\centering
\begin{tabular}{lllll}
\hline
\hline
Type                        & Injection                                                 &  Range   & Motion & Ways\\
of MHD turbulence  & velocity                                                   & of scales & type         & of study\\
\hline
Weak                       & $V_L<V_A$ & $[L_{inj}, l_{trans}]$          & wave-like & analytical\\
\hline
Strong                      &                      &                                        &                 &                \\
subAlfv\'{e}nic            &  $V_L<V_A$ & $[l_{trans}, l_{diss}]$ & eddy-like & numerical \\
\hline
Strong                    &                        &                                          &                 &                   \\
superAlfv\'{e}nic       & $V_L > V_A$ & $[l_A, l_{min}]$                    & eddy-like & numerical \\
\hline
& & & \\
\multicolumn{5}{l}{\footnotesize{$L_{inj}$ and $l_{diss}$ are injection and dissipation scales, respectively}}\\
\multicolumn{5}{l}{\footnotesize{$M_A\equiv u_L/V_A$, $l_{trans}=L_{inj}M_A^2$ for $M_A<1$ and $l_a=L_{inj}M_A^{-3}$ for $M_A>1$. }}\\
\end{tabular}
\end{table*}
}

{\toreferee  \subsection{Sub-Alfvenic Turbulence}} 
In the case of $V_L<V_A$, the Alfvenic Mach number is less than unity, i.e. $M_A=V_L/V_A < 1$. The turbulence in the range from the injection scale {\toreferee $L_{inj}$} to the transition scale 
\begin{equation}
l_{trans}=L_{inj}M_A^2
\label{trans}
\end{equation}
is termed the weak Alfvenic turbulence. This type of turbulence keeps the $l_{\|}$ scale stays the same while the velocities change as {\toreferee $v_\perp\approx V_L (l_\perp/L_{inj})^{1/2}$} (LV99, Gaultier et al. 2000). The cascading results in the change of the perpendicular scale of eddies {\toreferee $l_\perp$} only.  With the decrease of $l_\perp$ the turbulent velocities  {\toreferee $v_\perp$} decreases. Nevertheless, rather counter-intuitively, the strength of non-linear interactions of Alfvenic wave packets increases (see \citealt{L16} for the description of the interaction in terms of wave packets). Eventually, at the scale $l_{trans}$,  the turbulence transforms into the strong regime which obeys the GS95 critical balance. 

{\toreferee From the point of view of our gradient technique, the Alfvenic perturbations are perpendicular to the magnetic field. The slow modes (see below) are sheared by Alfvenic perturbations and also create gradients perpendicular to magnetic field. This statement is relevant both to velocity and density gradients.

The situations when the $l_{trans}$ is less than the turbulence dissipation scale $l_{diss}$ require $M_A$ that is unrealistically small for the typical ISM conditions. Therefore, typically the ISM turbulence transits to the strong regime. If the telescope resolution is enough to resolve scales less than $l_{trans}$ it is strong MHD turbulence that we make use of within our gradient technique.}

The anisotropy of the eddies for subAlfvenic turbulence is larger than in the case of trans-Alfvenic turbulence described by GS95. The following expression was derived in LV99:
\begin{equation}
\label{anis}
l_{\|}\approx L_{inj}\left(\frac{{\toreferee l_\perp}}{L_{inj}}\right)^{2/3} M_A^{-4/3}
\end{equation}
where $l_{\|}$ and $l\perp$ are given in the local system of reference. For $M_A=1$ one returns to the GS95 scaling.
The turbulent motions at scales less than $l_{trans}$ obey:
\begin{equation}
{\toreferee v_\perp}=V_L \left(\frac{{\toreferee l_\perp}}{L_{inj}}\right)^{1/3} M_A^{1/3},
\label{vel_strong}
\end{equation}
i.e. they demonstrate Kolmogorov-type cascade perpendicular to {\toreferee local} magnetic field. 

{\toreferee In the range of $[L_{inj}, l_{trans}]$ the direction of magnetic field is weakly perturbed and the local and global system of reference are identical. Therefore the velocity gradients calculated at scales larger than $l_{trans}$ are perpendicular to the large scale magnetic field.} {\toreferee While at} scales smaller than $l_{trans}$ the velocity gradients follow the direction of the local magnetic fields, similar to the case of trans-Alfvenic turbulence that we discuss in the main text.

{\toreferee  \subsection{Super-Alfvenic Turbulence}} If $V_L>V_A$, at large scales magnetic back-reaction is not important and up to the scale
\begin{equation}
l_A=L_{inj}M_A^{-3},
\label{la}
\end{equation}
the turbulent cascade is essentially hydrodynamic Kolmogorov cascade. At the scale $l_A$, the turbulence transfers to the {\toreferee sub}-Alfvenic turbulence described by GS95 scalings{\toreferee, i.e. anisotropy of turbulent eddies start to occur at scales smaller than $l_A$.}

The velocity gradients at the range from the injection scale $L_{inj}$ to $l_A$ are determined by hydrodynamic motions and therefore are not sensitive to magnetic field. The contribution from these scales is better to remove using spacial filtering. For scales less than $l_A$ the gradients reveal the local direction of magnetic field{\toreferee, as we described e.g. in  \cite{YL17b,Letal17} and also  discuss in the main text. For our numerical testing we are limited in the range of $M_A>1$ that we can employ. {\torefereetwo In the case when $M_A$ is sufficiently small, the scale $l_A$ will be comparable to the dissipation scale $l_{dis}$ and therefore the inertial range will be entirely eliminated. }Thus we have to limit our numerical testing to $M_A<2$. From the theoretical point of view, there are no limitations for tracing magnetic field within super-Alfvenic turbulence provided that the telescope or interferometer employed resolves scales less than $l_A$ and $l_A>l_{dis}$. 

\subsection{Cascades of fast and slow MHD modes}

In compressible turbulence, apart from Alfvenic motions, slow and fast fundamental motion modes are present (see Biskamp 2003). These are compressible modes and their basic properties are described e.g. in Brandenburg \& Lazarian (2013).  

In short, the three modes, Alfven, slow and fast have their own cascades (see Lithwick \& Goldreich 2001, Cho \& Lazarian 2002, 2003). Alfvenic eddy motions shear density perturbations coresponding to the slow modes and imprint their structure on the slow modes. Therefore the anisotropy of the slow modes mimic the anisotropy of Alfven modes, the fact that is confirmed by numerical simulations for both gas pressure and magnetic pressure dominated media (Cho \& Lazarian 2003, Kowal \& Lazarian 2010). Therefore the both velocity and magnetic field gradients are perpendicular to the local direction of magnetic field. This is confirmed in numerical testing in \citep{LY18a}. 

Fast modes for gas pressure dominated media are similar to the sound waves, while for the media dominated by magnetic pressure are waves corresponding to magnetic field compressions. In the latter case, the properties of the fast mode cascade were identified in Cho \& Lazarian (2002). The gradients arising from fast modes are different from those by Alfven and slow modes as shown in \citep{LY18a}. However, both theoretical considerations and numerical modeling (see Brandenburg \& Lazarian 2013) indicate the subdominance of the fast mode cascade compared to that of Alfven and slow modes. In addition, in realistic ISM at small scales fast modes are subject to higher damping (see Yan \& Lazarian 2004, Brunetti \& Lazarian 2007). In numerical simulations \citep{LY18a} the velocity gradients calculated with Alfvenic modes only were undistinguishable from those obtained with all 3 modes present. }

\section{Comparison with the correlation function anisotropy (CFA) technique}
\label{subsec:cfa}
The technique of measuring $M_A$ based on the correlation function anisotropy (henceforth CFA) was suggested as a technique to trace magnetic fields and find $M_A$ \cite{Letal02, EL05, Betal14, Eetal15}. The advantage of using the correlation function anisotropies is that there are analytical predictions connecting the expected anisotropies with the properties of slow, fast and Alfven modes \cite{KLP16, KLP17a, KLP17b}. The disadvantage of CFA compared to gradients is that, CFA require a larger statistical area for averaging compared to VGT (see \citealt{Letal17, YL17b}), thus CFA can only estimate a coarse structure of magnetic field. 

Revisited recently in \cite{YLL18} , it is possible to use {\toreferee the} CFA technique to trace {\toreferee the} magnetic field direction. The (second order) correlation function for Velocity Centroid is defined as:
\begin{align}
\label{eq:cf}
CF_{2;centroid}(\textbf{R})=\langle Ce(\textbf{r})Ce(\textbf{r}+\textbf{R})\rangle
\end{align}

The direction of the major axis of the correlation function determines how the {\it average} magnetic field in the region is {\toreferee oriented.} We would adopt the strategy as in \cite{brazil18} through Fast-Fourier Transform:
\begin{align}
\label{eq:fftc}
CF_{2;centroid}(\textbf{R})= \mathfrak{F}^{-1} \{|\mathfrak{F}\{Ce\}|^2\}
\end{align}
where $\mathfrak{F}$ is the Fourier transform operator. With the correlation function computed, we can then adopt the contour searching method as discussed in \cite{brazil18}. We here briefly explain their method using the cartoon from \cite{brazil18} as Fig \ref{fig:illuscfa} here.

Figure \ref{fig:illuscfa} shows how to locate the direction of anisotropy given a specific correlation map or structure function map. Concretely, the algorithm plots the contour lines of the map, and detect the orientation of the elongated major-axes and minor-axes of each (elliptical) contour line. Then the map is rotated such that the major-axis of the inner contour is parallel to the horizontal direction. The direction of anisotropy is then determined by the direction of major axis of the {\toreferee dark blue contour as shown in the middle panel of Fig \ref{fig:illuscfa}. In the right of Figure \ref{fig:illuscfa} shows the axis anisotropy plot for different contours. The mean direction of the major axis orientation determines the predicted magnetic field direction using the CFA technique. } 

\begin{figure*}[tbhp]
\centering
\includegraphics[width=0.96\textwidth]{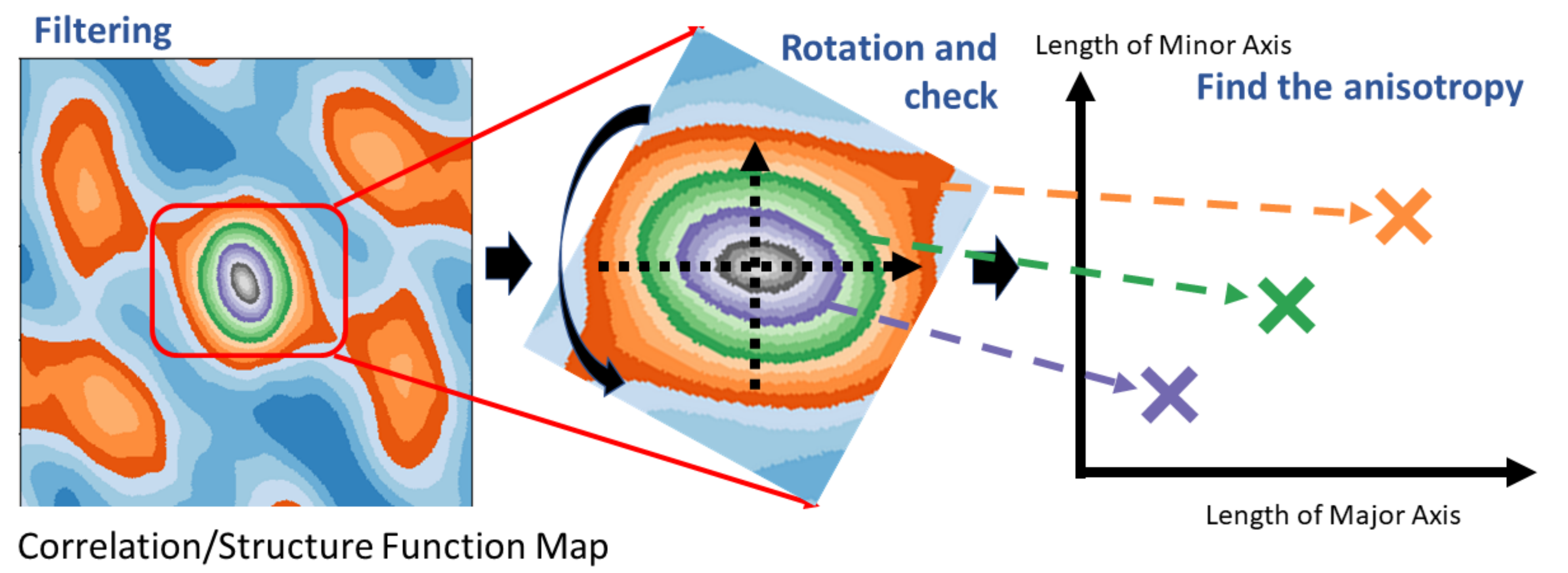}
\caption{\label{fig:illuscfa} An illustration showing how the direction of anisotropy is detected using the rotation-detection algorithm. (Left) We first locate the region having elliptical contours and put the rotation center on the origin of the ellipses. (Middle) Then we slowly rotate the contours so that we identify the major and minor axes {\toreferee and both axes length are recorded for different contours. For example,the big dashed arrows shows the major and minor axes of the dark blue elliptical structure}. (Right) The axes are then providing necessarily informations (direction, anisotropic length) for magnetic field studies. Captured from \cite{brazil18}.}
\end{figure*}

For our analysis, we compute the correlation function for the whole centroid map. In Figure \ref{fig2} we compare both $\sigma_{GD}$ and top-base ratio to the CFA axis ratio in a number of numerical cubes with different $M_A$. The CFA axis ratio has the similar trend with respect to $M_A$ compared to the top-base ratio and the variance. However, from \cite{brazil18} they showed that the method of CFA only works when the sampling area is around $100^2$ out of $792^2$, which limits the applicability of using CFA to obtain $M_A$, even though the method is well-studied \citep{EL05, Eetal15, KLP17a, KLP17b}. In contrast, the gradient technique, including the dispersions method in the present work, requires a smaller sampling area for a statistically significant result. The {\toreferee distribution} of gradient orientations parameters (variance, top-base ratio) therefore advantageous when observers want an estimation of magnetic field strength compared to the CFA method {\toreferee through substituting the dispersion of polarization angles in the Davis-Chandrasekhar-Fermi \citep{D51,CF53} technique to the dispersion of CFA orientations in the region of interest}. 

\section{Obtaining $M_A$ {\toreferee by} studying polarization degree}
\label{app}
To understand why there is {\toreferee a kink in Fig. \ref{fig3} \& \ref{fig4}} at around $M_A\sim 1$, we compare our result with the Chandrasekhar-Fermi (CF, see \citealt{CF53}) method with a slight modification : We compute the standard deviation of polarization percentage $\sigma_{pol\%}$ and compare with $M_A$ using a number of simulation cubes used in our previous studies \citep{YL17a,YL17b,LY18a}. From \cite{Fal08} we know that the traditional formulation of CF method does not work when $M_A$ is close to unity. Instead \cite{Fal08} suggested the a straightforward generalization of the traditional relation
$\delta \tan\theta_{pol} \propto M_A$, where  $\delta \theta_{pol}$ is the dispersion of the distribution of polarization angles.

In the situation that we sample magnetic field over line of sight scale ${\cal L}$ much larger than the turbulence injection scale. {\toreferee As a result,}  an additional factor {\toreferee is introduced in \cite{CY16} to compensate the contributions from multiple line-of-sight eddies in their modification of the Davis-Chandrasekhar-Fermi \citep{D51,CF53} technique:}
\begin{equation}
\alpha=\sqrt{\frac{L}{{\cal L}}},
\end{equation}
{\toreferee which} increases the relative importance of the observed mean magnetic field compared to the random magnetic field. Indeed, the integration of fluctuating magnetic field along the line of sight is a random walk process with the integral $\sim \sqrt{L{\cal L}}$, while the contribution of mean field increases as $\sim L$. Therefore the ratio of the $\int \delta B ds ~/\int B_0 ds\approx M_A \alpha$. This is the ratio that is available from observations. Thus we adopt:
\begin{equation}
\delta \tan\theta_{pol} \approx \delta \theta_{pol} \propto \alpha M_A,
\label{tan}
\end{equation}where we assumed small angle variations.
What is important for us is that $P_{\%}={\toreferee\sqrt{Q^2+U^2}/I}$ is proportional to {\toreferee the difference of cross sections} $C_x - C_y${\toreferee (See \citealt{LD85} for a detailed discussion), where $I,Q,U$ are the Stokes parameters.} The fluctuation of $P_\%$, i.e. $\delta P_\%=P_\%-\langle P_\% \rangle$, can be obtained assuming that fluctuations in angle is measured from the mean field direction. Thus for $\theta\ll 1$, $\delta P_\%\sim (\delta \theta_{pol})^2$ and the squared dispersion of polarization
\begin{equation}
\sigma_{pol\%}^2 = \langle (\delta P\%)^2 \rangle\sim (\delta \theta_{pol})^4,
\end{equation}
which means that
\begin{equation}
\sigma_{pol\%} \sim  M_A^2\alpha^2.
\end{equation}

Therefore  $\sigma_{pol\%} \sim \delta \theta_{pol}^2$ , we expect a $M_A^2$ dependence for small $M_A$. On the right Fig \ref{fig3}, we show the plot of $\sigma_{pol\%}-M_A$ using the simulations from LY18. The result in Fig. \ref{fig3} is equivalent to the modified CF method in F08.  

For our numerical study $\alpha \approx 1$ and we are not exploring the dependence\footnote{The dependence of $\sigma_{pol\%}$ on $\alpha$ is different of the dependence of the dispersion of velocity gradients on $\alpha$. The latter does not depend on the line of sight integration and therefore is independent of $\alpha$. Therefore, by finding the $M_A$ using the two approaches one can find the important physical parameter of the turbulent volume $L/{\cal L}$. } of $\sigma_{pol\%}$ on $\alpha$.  Instead, by using more than 30 numerical cubes with different $M_S$, $M_A$ and resolutions, Fig \ref{fig3} shows that there is a power-law relation of $\sigma_{pol\%} \propto M_A^2$ at $M_A<1$. The flattered fitting line with  with $M_A\in [1,2]$ has a proportionality of $\sigma_{pol\%} \propto M_A^{0.6}$. For gradient dispersion, the power law becomes  $(1-R)\propto M_A^{0.06}$. The ratio of power-law index and also the behavior in the two regimes of $M_A$ are also similar for both gradient dispersion method and CF method. The above analysis suffices to show that the gradient dispersion technique can estimate the magnetization of a system in a way similar to the CF method.

\end{document}